\documentclass{emulateapj}
\usepackage{times}
\usepackage{epsfig}
\usepackage{hyperref}

\newcommand{\tnm}{\tablenotemark}
\newcommand{\tnt}{\tablenotetext}

\newcommand{\ergs}{erg s$^{-1}$}

\newcommand{\average}[1]{\ensuremath{\langle#1\rangle} }

\newcommand{\msun}{\ensuremath{M_{\sun}}}

\newcommand{\bootes}{Bo\"{o}tes} 
 
\newcommand{\spitzer}{{\it Spitzer}}

\newcommand {\apgt} {\ {\raise-.5ex\hbox{$\buildrel>\over\sim$}}\ }
\newcommand {\aplt} {\ {\raise-.5ex\hbox{$\buildrel<\over\sim$}}\ }

\newcommand{\ben}{\begin{enumerate}}
\newcommand{\een}{\end{enumerate}}
\newcommand{\bit}{\begin{itemize}}
\newcommand{\eit}{\end{itemize}}
\newcommand{\beq}{\begin{equation}}
\newcommand{\eeq}{\begin{equation}}

\def\gtrsim{\mathrel{\hbox{\rlap{\hbox{\lower4pt\hbox{$\sim$}}}\hbox{\raise2pt\hbox{$>$}}}}}

\newcommand{\loiii}{\ensuremath{L_{\mathrm{[O {\tiny III}]}}}}

\newcommand{\oiii}{[\ion{O}{3}]}

\newcommand{\threefour}{\ensuremath{{\rm W1}-{\rm W2}}}
\newcommand{\fourtwelve}{\ensuremath{{\rm W2}-{\rm W3}}}
\newcommand{\wise}{{\em WISE}}
\newcommand{\galex}{{\em GALEX}}

\newcommand{\aten}{\citetalias{asse10agntemp}}
\newcommand{\rsix}{\citetalias{rich06}}

\def\lax{{$\mathrel{\hbox{\rlap{\hbox{\lower4pt\hbox{$\sim$}}}\hbox{$<$}}}$}}
\def\gax{{$\mathrel{\hbox{\rlap{\hbox{\lower4pt\hbox{$\sim$}}}\hbox{$>$}}}$}}

\shorttitle{{\it SDSS and {\wise} quasar SEDs and colors}}
\shortauthors{Hickox et al.}
\slugcomment{Accepted for publication in The Astrophysical Journal}

\begin{document}

\title{Composite spectral energy distributions and
infrared--optical\\ colors of type 1 and type 2 quasars}

\shortauthors{HICKOX ET AL.}
\author{
Ryan C.\ Hickox\altaffilmark{1}, 
Adam D.\ Myers\altaffilmark{2,3},
Jenny E.\ Greene\altaffilmark{4},
Kevin N.\ Hainline\altaffilmark{5},
Nadia L.\ Zakamska\altaffilmark{6},\\
Michael A.\ DiPompeo\altaffilmark{1}
}

\altaffiltext{1}{Department of Physics and Astronomy, Dartmouth College, 6127 Wilder Laboratory, Hanover, NH 03755, USA; ryan.c.hickox@dartmouth.edu}
\altaffiltext{2}{Department of Physics and Astronomy, University of Wyoming, Laramie, WY 82071, USA}
\altaffiltext{3}{Max-Planck-Institut f\"ur Astronomie, K\"onigstuhl 17, D-69117  Heidelberg, Germany}
\altaffiltext{4}{Department of Astrophysics, Princeton University, Princeton, NJ 08544-1001, USA}
\altaffiltext{5}{Steward Observatory, University of Arizona, 933 North Cherry Avenue, Tucson, AZ 85721, USA}
\altaffiltext{6}{Department of Physics and Astronomy, Johns Hopkins University, Bloomberg Center, 3400 North Charles Street, Baltimore, MD 21218, USA}

\begin{abstract}

We present observed mid-infrared and optical colors and composite spectral energy
distributions (SEDs) of type 1 (broad-line) and 2 (narrow-line) quasars selected from Sloan
Digital Sky Survey (SDSS) spectroscopy. A significant fraction of powerful quasars
are obscured by dust, and are difficult to
detect in optical photometric or spectroscopic surveys. However these may be more easily identified on the basis of mid-infrared (MIR) colors and SEDs. Using samples of
SDSS type 1 type 2 matched in redshift and \oiii\ luminosity, we
produce composite rest-frame 0.2--15 \micron\ SEDs based on SDSS, UKIDSS, and {\em Wide-Field Infrared Survey Explorer} ({\wise})
photometry and perform model fits using simple galaxy and quasar SED
templates. The SEDs of type 1 and 2 quasars are remarkably similar,
with the differences explained primarily by the extinction of the
quasar component in the type 2 systems. For both types of quasar, the flux of the AGN relative to the host galaxy increases with AGN luminosity (\loiii) and redder observed MIR color, but we find only weak dependencies of the composite SEDs on mechanical jet power as determined through radio luminosity.   We conclude that luminous quasars can be effectively
selected using simple MIR color criteria similar to those identified previously ($\threefour > 0.7$; Vega), although these criteria miss many heavily obscured objects. Obscured quasars can be further identified based on optical-IR colors (for example, $(u-{\rm W3\;[AB]})>1.4({\rm W1-W2\;[Vega]})+3.2$). These
results illustrate the power of large statistical studies of obscured
quasars selected on the basis of mid-IR and optical photometry.

\end{abstract}


\section{Introduction}

Studies of optically luminous quasars have yielded remarkable insights 
into the growth of supermassive black holes (BHs) over cosmic time (see \citealt{alex12bh} for a review). 
Analyses of the broad-band spectral energy distributions (SEDs) of 
unobscured (``type 1'') quasars have elucidated the physics of BH accretion
(\citealt{elvi94, rich06, kell10qsoedd}), and surveys in the soft X-rays
and optical have shown that quasar activity (and thus BH growth) peaks
at early cosmic times ($z\sim 2$--3; e.g., \citealt{croo04twodfqz, rich05, hasi05, 
fan06}). Quasar clustering measurements \citep[e.g.,][]{porc04clust,croo05,
coil07a, myer07clust1,shen07clust, daan08clust,padm09qsored, 
ross09qsoclust, krum10xclust, dipo16qsoclust, dipo17qsoclust} suggest that the processes that fuel 
rapid BH growth are tied to the buildup of large-scale structure 
\citep[e.g.,][]{hopk08frame1, crot09qsohalo,conr13qso} and indicate that quasars may play a 
role in regulating star formation and in the emergence of the red galaxy 
population \citep[e.g.,][]{coil08galclust,brow08halo,tink10clust, thac14agnsf, voge14illustris, scha15eagle}.

These many successes reveal only part of the story. It has long been
known that BH growth can occur behind large columns of gas and dust
\citep[e.g.,][]{sett89, coma95}.  Such obscured (``type 2") active
galactic nuclei (AGN) can be identified from narrow optical emission
lines \citep{zaka03,zaka04,zaka05, reye08qso2,yuan16qso2}, radio luminosity
\citep[e.g.,][]{mcca93highzradio, mart06, seym07radiohosts, wilk13radiox}, or X-ray
properties \citep[e.g][] {alex01xfaint, ster02, trei04, vign06qso2,
  vign09qso2, delm14nustar}. In addition, pioneering work with the {\em Spitzer
  Space Telescope} demonstrated that obscured quasars have similar
mid-infrared (MIR) SEDs to their unobscured counterparts, but are dominated by
host galaxy light in the optical \citep{lacy04, ster05, rowa05,
  poll06, alon06, mart06, hick07abs, donl08spitz, lacy13spec, lacy15iragn, dai14qso}. Consequently, obscured
quasars can be most efficiently selected in the MIR, as they appear
very red in the IRAC [3.6]--[4.5] color, characteristic of the ``hot'' MIR
SED that is evident in broad-line quasars \citep[e.g.,][hereafter R06 and A10, respectively]{rich06, asse10agntemp}. However, in contrast to their unobscured counterparts, obscured quasars have very red optical-IR colors, due to extinction of the nuclear emission in the rest-frame optical and UV \citep[e.g.,][]{poll06, hick07abs}. {\em Spitzer} therefore unveiled significant samples of
obscured quasars \citep[see \S1 of][]{hick11qsoclust} and MIR
studies find roughly equal numbers of obscured and unobscured quasars
\citep[e.g.,][]{hick07abs, asse15wiseqso}.

\defcitealias{rich06}{R06}
\defcitealias{asse10agntemp}{A10}

Obscured quasars thus represent a large fraction of the massive BH
growth in the Universe, but it remains unclear if the obscuration is due
to a dusty torus intrinsic to the central engine (as posited in the
``unified model'' of AGNs; \citealt[e.g.,][]{anto93, urry95}), to
larger-scale clouds in the host galaxy \citep[e.g.,][]{sand88, page04submm,
goul12comp, chen15qsosf}, or to material distributed over a wide range of scales. It is
also unclear whether obscured quasars eventually evolve into unobscured
objects as they blow away their surrounding dust, as posited by some
evolutionary models \citep[e.g.,][]{sand88,dima05qso,hopk08frame1}. To
answer these questions, we require significant statistical samples of
obscured quasars, but samples selected with {\em Spitzer} are small,
numbering $\sim$1000 at most relative to $>10^6$ known unobscured
quasars \citep[e.g.,][]{rich09qsophot}. Thus our understanding of
obscured quasars as a cosmological population remained comparatively limited. 

The completion of the {\em Wide-Field Infrared Survey Explorer}
\citep[{\em WISE};][]{wrig10wise} survey of the full sky has allowed us to make
dramatic progress in statistical studies of AGN \citep[e.g.,
][]{mor11wise, plot12wise, dono12wise, edel12wise, ster12wise,
eise12wise, ichi12iragn, saji12iragn, mate12xmmwise, asse13wiseagn, geac13qsocmb, dipo14qsoclust,  dipo15qsocmb, dipo16qsoclust, dipo17qsoclust,
secr15wise, asse15wiseqso, asse17wise}. In particular, {\wise} enables us
to detect and characterize hundreds of thousands of obscured quasars,
increasing the sizes of obscured quasar samples by two orders of
magnitude and allowing the first large statistical studies of their
properties.

In this paper we use {\wise} photometry to explore the MIR colors and
composite broad-band SEDs of obscured and unobscured quasars selected
from the Sloan Digital Sky Survey (SDSS), using photometry from {\em WISE}, SDSS, the UKIRT Infrared Deep Sky Survey (UKIDSS), the Two-Micron All-Sky Survey (2MASS), and the {\em Galaxy Evolution Explorer} (\galex{}). This work builds on several
recent studies that have developed various AGN color selection
criteria using {\wise} photometry \citep[e.g.,][]{jarr11wise, ster12wise, 
  mate12xmmwise}. Here we focus specifically on rare, luminous
obscured quasars identified using optical spectroscopy, for which the largest sample consists of the type 2 quasars from SDSS \citep{zaka03,
  reye08qso2, yuan16qso2}. We compute the optical and MIR colors and composite
SED of these obscured quasars, and compare them to optically-selected
unobscured quasars that are matched in redshift and luminosity, with
the aim of developing photometric criteria that will allow us to
select large samples of very luminous obscured and unobscured
quasars for future measurements of space densities, clustering, and
other statistical studies. We further explore the dependence of quasar SED shapes on physical and observational properties (AGN luminosity, MIR color, and radio-loudness).
The paper is organized as follows: In
\S~\ref{sec:sample} and \S~\ref{sec:photo} we introduce our SDSS
quasar samples and the SDSS and {\wise} photometric data set, in
\S~\ref{sec:seds} we compute composite optical-IR SEDs for the two
sets of quasars, and in \S~\ref{sec:colors} we explore photometric
selection criteria. In \S~\ref{sec:discussion} we summarize our
results and discuss future applications of these results. Throughout the paper we assume a $\Lambda$CDM cosmology with $H_0 = 70$ km s$^{-1}$ Mpc$^{-1}$, $\Omega_m = 0.3$, and $\Omega_\Lambda=0.7$.

\section{Quasar samples}

\label{sec:sample}

\begin{figure}
\epsscale{1.15}
\plotone{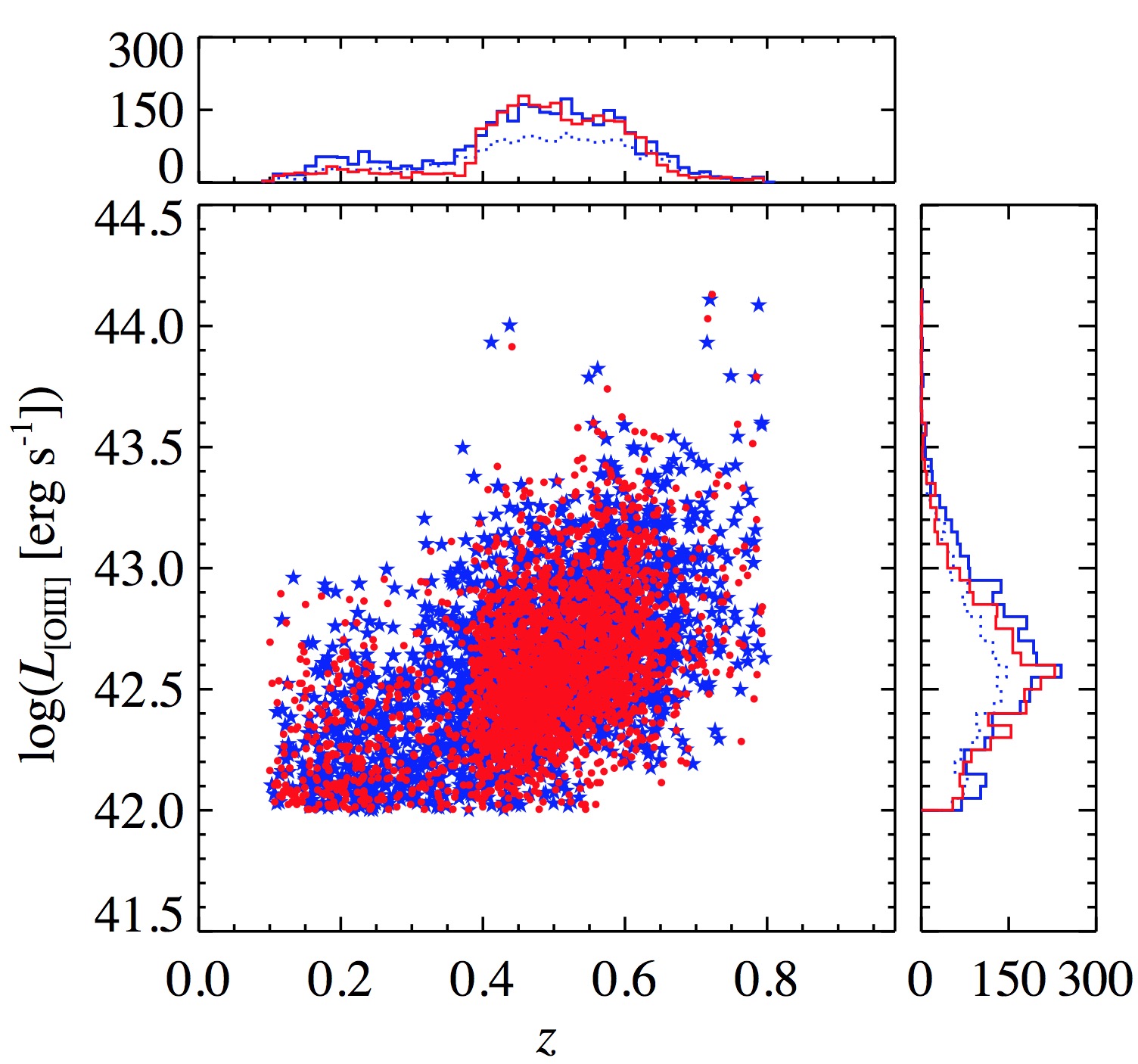}
\caption{\label{fig:lz} Distribution in redshift and \ion{O}{3}
  luminosity for the sample of SDSS type 2 quasars (QSO 2s; red
  circles) and the sample of type 1 quasars (QSO 1s; blue stars)
  matched in $z$ and \loiii. The dotted blue histograms show the distributions for the full sample of QSO 1s, while the solid blue histograms show the distribution accounting for QSO 1s that were matched in $z$ and \loiii\ to more than one QSO2, as described in \S~\ref{sec:sample}. }
\end{figure}

We employ two samples of quasars selected from the spectroscopic
database of SDSS. The first sample (QSO 1s) consists of broad-line
(type 1) quasars selected for spectroscopy based on their blue optical-to-UV
colors. These objects are spectroscopically confirmed to have broad
permitted emission lines indicative of gas moving in the gravitational
potential of a black hole.  The other sample consists of obscured,
narrow-line (type 2) QSOs. These were targeted by SDSS based on unusual colors or detections at X-ray or radio wavelengths, but were then spectroscopically selected to
have luminous emission in the \oiii$\lambda$5007 \AA\ emission line.

Broad-line objects were selected from the SDSS catalog
of \citet{shen11sdssqso}.  This work presents spectroscopic
measurements for a well-defined sample of optical-UV selected
unobscured QSOs from the Seventh Data Release (DR7) of the SDSS
\citep{abaz09sdss}.  As mentioned above, approximately half of the sources
are selected for observation originally because of their blue colors
\citep{rich02sdss}.  The remainder were either targeted by
earlier versions of the QSO selection algorithm or for some
`serendipitous' reason (such as radio or X-ray detection).  For our
purposes we do not distinguish between sources based on targeting.
\citeauthor{shen11sdssqso} fit the quasar continuum (including the
broad \ion{Fe}{2} pseudocontinuum), and model the broad and narrow
lines with multi-Gaussian fits.  From the \citeauthor{shen11sdssqso}
catalog we extract the $\sim 19,000$ QSOs that fall in the redshift
range of interest $0.1 < z < 0.8$ (chosen to match the approximate
redshift range of the obscured QSOs described below and to exclude
contamination at very low redshift). 

As discussed
below, we use the \oiii\ luminosity (\loiii; a proxy for bolometric AGN luminosity; e.g., \citealt{lama10lbol}) 
to produce matched samples of obscured and unobscured sources. We use \loiii\ as our primary AGN luminosity indicator because it readily available for all sources, can be used for both type 1 and 2 quasars, and was produced directly from the spectroscopic analysis from which the samples were selected. However, we note that \loiii\ is not a perfect indicator of intrinsic AGN power. For example, \loiii\ is found to correlate only weakly with intrinsic hard X-ray luminosity, which is a more direct measure of the instantaneous power of the central engine \citep{bern15batagn}. Further, it is possible for the \oiii\ flux to be affected by dust extinction in the host galaxy \citep[e.g.,][]{zaka06host}, and the relationship between \loiii\ and bolometric luminosity may be complex, depending for example on the extent of the emitting narrow-line region \citep[e.g.,][]{huse13nlr, hain13salt, hain14nlr, hain16mdm}.

The type 2 quasars (QSO 2s) were taken from \citet{reye08qso2} and \citet{yuan16qso2}, which
expand the original search of
\citet{zaka03}. \citeauthor{reye08qso2} and \citeauthor{yuan16qso2} begin with the SDSS
spectroscopic database, and select targets with emission line fluxes
indicative of photoionization by an AGN, either through diagnostic line ratios \citep[e.g.,][]{bald81bpt, kewl01opt} or through the equivalent widths of the \oiii\ or [\ion{Ne}{5}]$\lambda$3426 \AA\ emission lines \citep[e.g.,][]{zaka03, gill10nev}.  They remove Type 1 (unobscured) QSOs by eliminating sources with
broad permitted emission (typically H$\beta$) with line widths that are substantially broader than that observed for \oiii\ as determined through visual inspection of the line fits. For their high-redshift ($z>0.52$) sample, \citeauthor{yuan16qso2} applied an initial limit of ${\rm FWHM(H\beta)}<1000$ km s$^{-1}$ to remove broad-line sources; visual inspection of the spectra identified a small number (10) of QSO 2s with ${\rm FWHM(H\beta)}>1000$ km s$^{-1}$. After compiling the full QSO 2 catalogs from the \citeauthor{reye08qso2} and \citeauthor{yuan16qso2} catalogs, we follow \citeauthor{reye08qso2} and
impose a luminosity cut of \loiii$>10^{8.3}~L_{\sun}$ for completeness.  Unlike the broad-line quasars, which were specifically
targeted by the SDSS based on color, the obscured sources were
targeted for a variety of reasons, primarily because many were included in spectroscopic follow-up by the serendipitous targeting algorithm.  Some had the colors of high
redshift QSOs because of the high equivalent width emission lines,
while some had radio or X-ray counterparts; we refer the reader to \citet{reye08qso2} and \citet{yuan16qso2} for the full details of the sample selection.

To enable a statistically useful comparison between the unobscured and
obscured quasar populations, we define a subset of the
full type 1 quasar sample that is selected have the same distribution as the QSO 2s in bins
of width 0.07 in redshift and 0.1 in $\log{\loiii}$. This selection yields
3425 "matched" type 1 quasars. We note that some QSO 1s are matched to multiple QSO 2s (approximately 20\% of the QSO 1s for our main SDSS+UKIDSS+\wise{} analysis); for these objects the matched QSO 1s are included multiple times in the averaging of the SEDs, so best to mirror the $z$ and $\loiii$ distributions of the QSO 2s.
This matched comparison sample of type 1 quasars will comprise the sample referred to as QSO
1s for the remainder of the paper. The distributions in $z$ and
\loiii\ for the QSO 1s and 2s shown in Figure~\ref{fig:lz}.

\section{Photometric data}

\label{sec:photo}

Photometric data for studying SEDs and colors are taken from the SDSS Data Release 9 (DR9; \citealt{ahn12sdss}), UKIDSS Large Area Survey (LAS; \citealt{lawr07ukidss}), and the AllWISE catalog \citep{wrig10wise}. As a cross-check on our SEDs for a limited subset of bright objects, we also make use of UV photometry from the \galex{}{} All-sky Imaging Survey (AIS; \citealt{bian14galex}) and near-IR (NIR) photometry from the Two Micron All Sky Survey (2MASS; \citealt{skru06}).

For {\em ugriz} photometry we use SDSS model
magnitudes (in the AB system), corrected for Galactic reddening, as obtained from
the {\tt SpecPhoto} database; the 95\% completeness limits for SDSS photometry are 22.0, 22.2, 22.2, 21.3, and 20.5 (AB) for the  {\em u}, {\em g}, {\em r}, {\em i}, and {\em z} bands, respectively. Essentially all ($>$99.7\%) of the quasars are detected at $>2\sigma$ significance in each of the SDSS bands, except the $u$ band, for which only 69\% of the QSO 2s (but $>$99.7\% of the QSO 1s) have a significant detection.

{\wise} photometric data were obtained using the Gator online database to
search for all objects in the AllWISE Source Catalog at the
positions of the SDSS quasars. While the AllWISE flux limits vary depending on position on the sky, typical magnitude limits at high Galactic latitude are $\sim$17, 16, 12, and 9 (Vega) in the 3.4, 4.6, 12, and 22 \micron\ bands (W1--W4, respectively). Of the 3425 QSO 1s and 2892 QSO 2s, all
but 19 (99.4\%) and 143 (95.1\%) respectively have {\wise} counterparts within 2\arcsec. We use observed 
magnitudes in each of the W1--W4 bands, and convert to
fluxes using standard conversions from Vega to AB
magnitudes\footnotemark. Among the QSO 1s, 99\% and 93\% have
$>$2$\sigma$ detections in the W3 and W4 bands,
respectively, while the corresponding fractions for QSO 2s are 87\%
and 76\%.

\setcounter{footnote}{6}
\footnotetext{Magnitude offsets, where $\Delta m = m_{\rm AB}-m_{\rm Vega}$, are 2.699, 3.339, 5.174, and 6.620 in the W1--W4 bands, respectively.}

For NIR photometry, we obtained UKIDSS LAS {\em Y}, {\em J}, {\em H}, and {\em K} band catalogs from the LAS DR9 database \citep{lawr12ukidssdr9}, which has a 5-$\sigma$ depth for point sources of $K\approx18.4$ (Vega). We match to the SDSS positions within 1 arcsec. We use Petrosian magnitudes and convert to fluxes using the standard conversion from Vega to AB magnitudes \citep{hewe06ukirt}. Of the 3425 QSO 1s and 2892 QSO 2s, 1052 and 860 lie within the 4000 deg$^2$ UKIDSS LAS footprint, and of these 971 (94\%) and 754 (90\%), respectively, have counterparts within 1\arcsec{} that have significant detections in all four UKIDSS bands.

Given these high detection fractions in the coverage areas, we focus our primary SED analysis on the quasars (both type 1 and 2) having SDSS, {\wise}, and UKIDSS detections. However we also utilize the full SDSS+{\wise} sample for studies of the color distributions. In general, {\wise} colors are presented
in the Vega system, while colors involving optical magnitudes are
presented in the AB system. To construct SEDs we compute fluxes at the central wavelengths of the bands for SDSS {\em ugriz} (0.36,
0.47, 0.62, 0.75, and 0.89 \micron), UKIDSS {\em Y}, {\em J}, {\em H}, and {\em K} (1.03, 1.25, 1.63, and 2.20 \micron), and {\wise} W1--W4 (3.4, 4.6, 12, and
22 \micron).

As a cross-check of our analysis, we also utilize \galex{} and 2MASS photometry for a subset of brighter, lower-redshift sources. \galex{} AIS sources were taken from the catalog of \citet{bian11galex}, which reaches $5\sigma$ depths of $\approx$19.9 and $\approx$20.8 (AB) in the FUV and NUV bands, respectively. We find that 2558  QSO 1s and 1919 QSO 2s lie within the AIS footprint (within 1\arcmin\ of a \galex{} source), and of these, 2549 (96\%) and 713 (37\%), of the QSO 1s and 2s, respectively are matched to the \galex{} positions within 3\arcsec. \galex{} fluxes are obtained from the \citeauthor{bian11galex} catalog, and corrected for Galactic reddening following \citet{bian11galexsf}. We obtain 2MASS photometry from the Point Source Catalog \citep{cutr03twomasspsc}, with a depth of $K_s\approx 15$ (Vega), and Extended Source Catalog \citep{skru06twomassxsc}, with a depth of $K_s\approx 14$ (Vega); 1654 (48\%) and 300 (10\%) of the QSO 1s and 2s, respectively, are matched to 2MASS sources within 1.5\arcsec.  We convert 2MASS magnitudes to monochromatic fluxes following \citet{cohe03twomass}. The central wavelengths for the \galex{} FUV and NUV and 2MASS {\em J}, {\em H}, and $K_s$ bands are 0.15, 0.23, 1.24, 1.66, and 2.16 \micron, respectively.  The \galex{} and 2MASS detection fractions, particularly for the QSO 2s, are a strong function of redshift and rise to $>$99\% for QSO 1s and 93\% for QSO 2s for both 2MASS and \galex{} if we limit the redshift range to $z<0.2$. We therefore focus our \galex{} and 2MASS analysis to quasars at $z < 0.2$.

Finally, we utilize radio data to distinguish between quasars that are radio-loud (RL) and radio-quiet (RQ), corresponding to the presence or absence of strong relativistic jets \citep[e.g.,][]{pado16radio, pado17radio}. We perform a cross-match of the QSO samples to the Faint Images of the Radio Sky at Twenty-Centimeters (FIRST; \citealt{helf15first}) and NRAO VLA Sky Survey (NVSS; \citealt{cond98nvss}) catalogs, with point source detection sensitivities at 20 cm of $\approx$1 mJy and $\approx$2.5 mJy, respectively. We search for radio counterparts to the QSOs within 1.5\arcsec\ for the FIRST catalog and 10\arcsec\ for the NVSS catalog; these radii reflect the different angular resolutions of the two surveys ($\approx$5\arcsec\ and 45\arcsec, respectively) and are chosen to be the angle at which the frequency of matches approximately reaches that of the background at large separations (the results are insensitive to the precise choice of match radius). We identify a radio counterpart for 723 (21\%) of QSO 1s and 1075 (37\%) of QSO 2s. For the 45\% of radio counterparts with both FIRST and NVSS detections, most show good agreement in the fluxes. For $\approx$10\% of cases the NVSS flux is higher by greater than a factor of 2, likely corresponding to additional flux resolved within the larger NVSS beam. For sources with both FIRST and NVSS detections, we adopt the higher value of the flux for the subsequent analysis (broadly following the approach of \citealt{zaka04}).

For each radio source we calculate the radio power $P_{1.4\;
\rm{GHz}}$ (in W Hz$^{-1}$), performing a small $K$-correction assuming the
typical spectrum of faint 1.4 GHz sources \citep[$\alpha\approx0.5$,
where $S_\nu\propto \nu^{-\alpha}$;][]{pran06radio}. We select quasars as RL based on a monochromatic 1.4 GHz luminosity threshold of $10^{24}$ W Hz$^{-1}$. This is the approximate luminosity limit for the FIRST survey at $z=0.8$, is characteristic of moderately powerful radio AGN  \citep[e.g.,][]{hick09corr, tadh16radio}, and is above the luminosity for typical star-forming galaxies at low redshift \citep{kauf08radio}. Among our samples, 456 (13\%) of the QSO 1s and 600 (21\%) of the QSO 2s are identified as RL. Despite the difference in detection fractions,  \citet{zaka04} demonstrate that after accounting for the selection bias toward more radio-bright sources, the intrinsic RL fraction of optically selected type 2 quasars is similar to that of optically selected type 1s. We note that some QSOs, and particularly QSO 2s, are expected to have lobes that extend up to Mpc in scale, well beyond our angular matching radius. Thus sources that have only bright lobe emission with no compact core may be missed by our matching procedure. However, \citet{zaka04} performed a careful matching analysis of the \citet{zaka03} SDSS QSO2 sample to the FIRST catalog, including a visual examination of the FIRST images for those sources that did not have a counterpart within 3\arcsec. \citet{zaka04} found that only a few percent of the QSO 2s had extended emission in the FIRST images but were not matched to a compact core. This small fraction of missed sources indicates that this incompleteness will have a minimal effect on our analysis, which focuses on composite SEDs for the full RL and RQ samples.

\section{Composite Quasar SEDs}

\begin{figure}
\epsscale{1.15}
\plotone{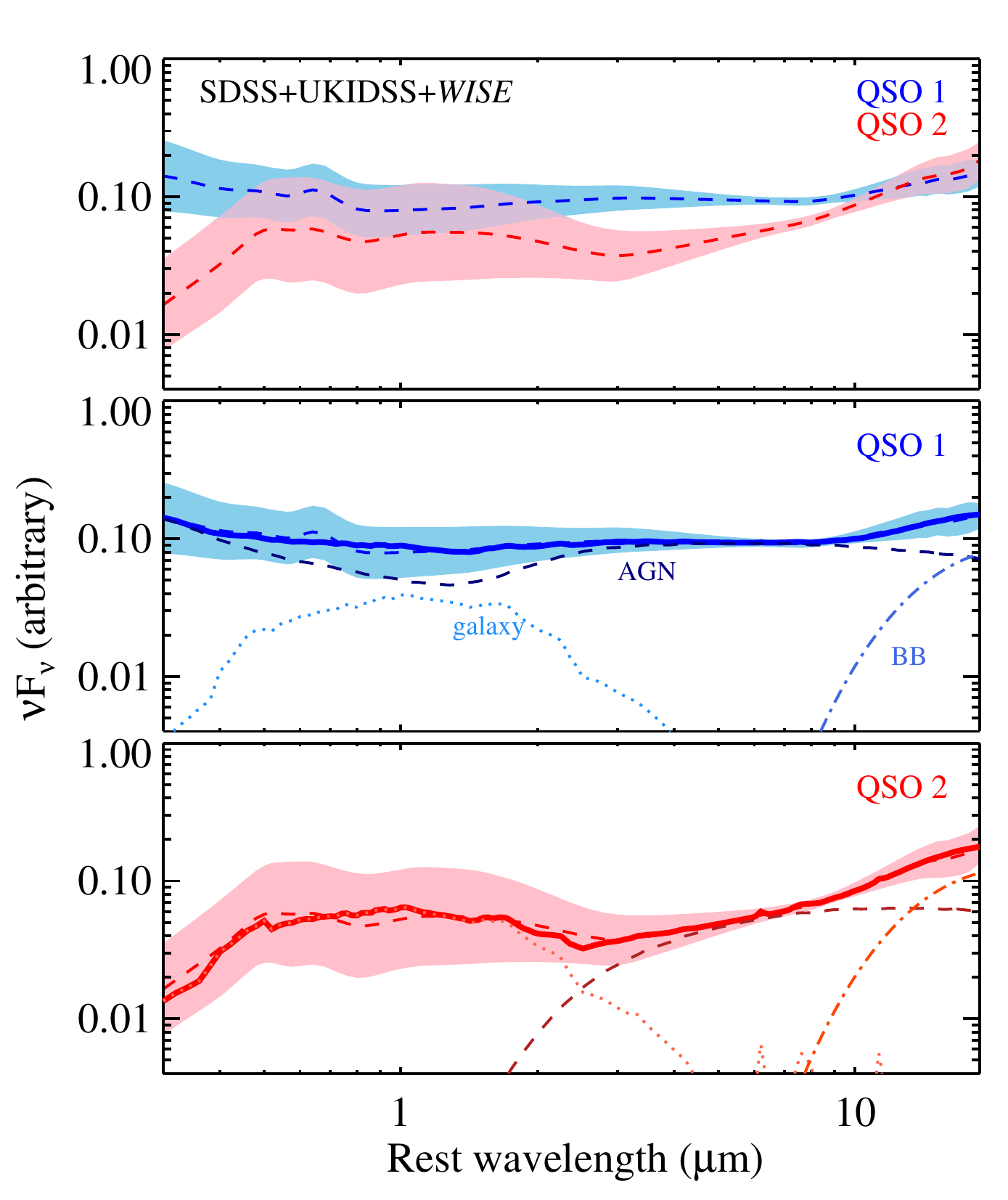}
\caption{The top panel shows the normalized composite SEDs for the QSO
  1s (blue) and QSO 2s (red), computed as described in
  \S~\ref{sec:seds}. Each individual quasar SED is normalized by the integrated flux
  between 8 and 13 \micron\ and averaged to create the composite; these are then scaled according to the relative flux of their (unabsorbed) AGN components derived from the SED fitting procedure described in \S~\ref{sec:seds}. The middle and bottom panels show the fits to
  the composite SEDs using a three-component model including host galaxy (dotted lines),
  AGN (dashed lines), and blackbody (dot-dashed lines) components. Here the dotted line shows the sum of the E and Im galaxy components. Note that the two SEDs show 
  similar ratios of AGN and host galaxy flux, with the primary
  difference being the reddening of the quasar continuum in the QSO
  2s. These best-fit model components are used to make predictions for
  the observed colors of obscured quasars as a function of redshift, as
  described in \S~\ref{sec:colors}. \label{fig:seds} }
\end{figure}

\begin{deluxetable}{ccccc}
\tablecaption{Composite QSO 1 and 2 SEDS \label{tab:seds}}
\tablehead{
\colhead{} &
\multicolumn{2}{c}{Scaled QSO 1 SED} &
\multicolumn{2}{c}{Scaled QSO 2 SED} \\
\colhead{$\log_{10}(\lambda \; {\rm [\mu m]})$} &
 \colhead{$\average{\log_{\rm 10}(F_\nu)}$} &
 \colhead{Scatter (dex)} &
  \colhead{$\average{\log_{\rm 10}(F_\nu)}$} &
 \colhead{Scatter (dex)} 
 }
\startdata
-0.650 & -1.420 &  0.254 & -2.668 &  0.277\\
-0.625 & -1.445 &  0.269 & -2.594 &  0.320\\
-0.600 & -1.436 &  0.272 & -2.510 &  0.327\\
-0.575 & -1.417 &  0.271 & -2.439 &  0.326\\
-0.550 & -1.395 &  0.266 & -2.386 &  0.328
\enddata
\tablecomments{The composite SEDs presented here are for the primary sample using SDSS, UKIDSS, and \wise{} photometry,  produced as described in Section~\ref{sec:seds} and as shown in Figure~\ref{fig:seds}. Table 1 is published in its entirety in the machine-readable format, and is available at \url{http://www.dartmouth.edu/~hickox/Hickox2017_QSO_SED_Table1.txt}
      A portion is shown here for guidance regarding its form and content.}
\end{deluxetable}

\begin{figure}
\epsscale{1.15}
\plotone{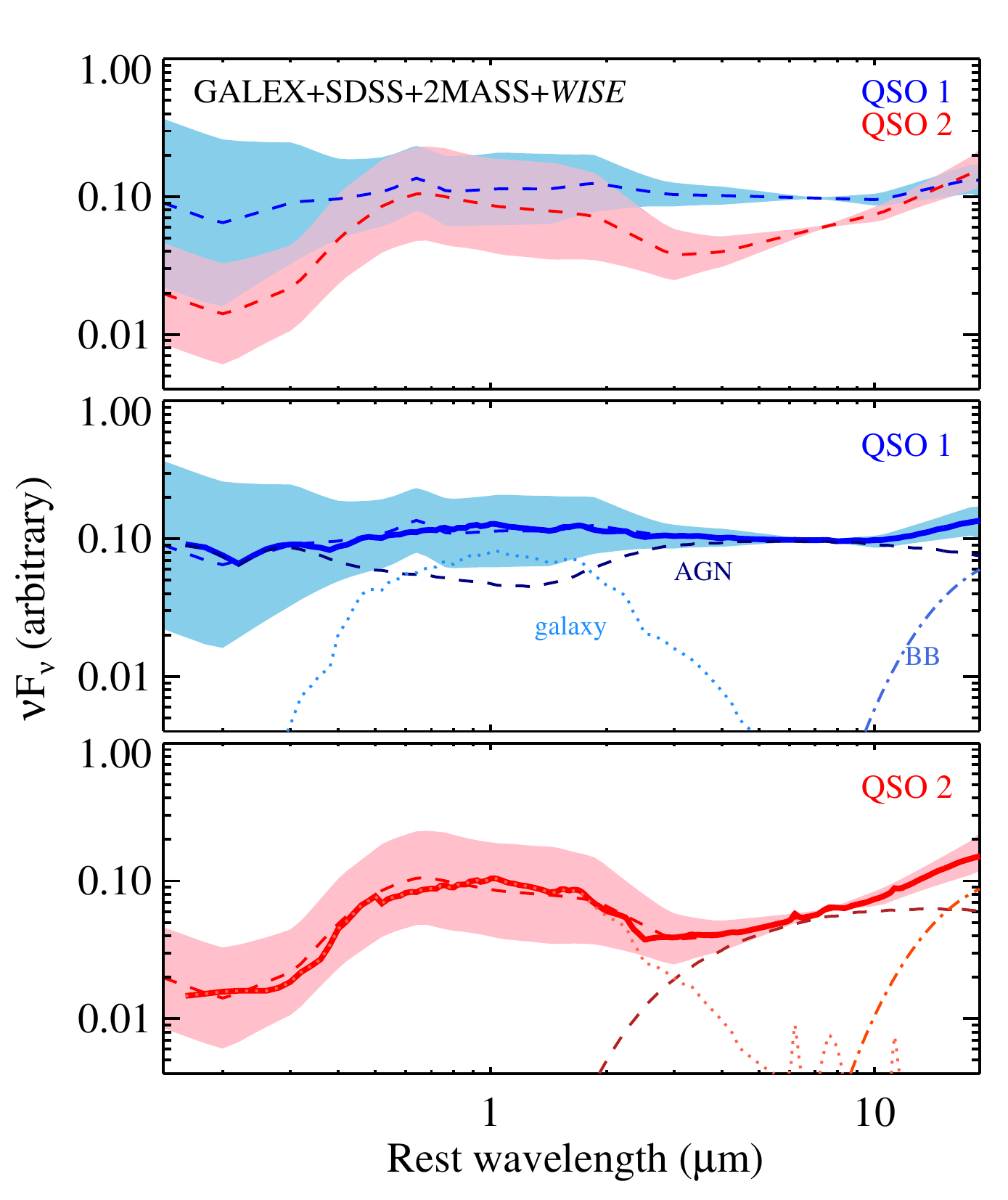}
\caption{Composite QSO 1 and 2 SEDs and model fits, as shown in Figure~\ref{fig:seds}, but  using 2MASS NIR data as well as UV photometry from \galex{}. The quasar samples are limited to sources at $z<0.2$, for which the completeness of 2MASS and \galex{} detections are $>93$\%. These composite SEDs and model fit parameters are similar to those determined using UKIDSS NIR photometry over the redshift range of the full sample, the primary difference being a somewhat smaller contribution from the AGN. \label{fig:twogalseds} }
\end{figure}

\label{sec:seds}

We next derive composite SEDs of the QSO 1 and QSO 2 samples. The
composite SEDs allow us to compare the overall physical characteristics of
the populations, and to model how the selection of these objects will
depend on redshift, luminosity, and other parameters.

We use a simple procedure to construct a mean SED for each of the two
populations. We include only sources with detections in all 12 SDSS, UKIDSS, and
{\wise} bands, as described in \S~\ref{sec:photo}, which represents 334 QSO 1s and 406 QSO 2s, respectively.  Because of the relatively high detection rates in each band (\S~\ref{sec:photo}), these criteria are not expected to significantly bias the results toward or away from any particular subsample. We confirm this explicitly by performing the same analysis described here, but including sources with marginal detections for which fluxes are given in the SDSS and {\wise} catalogs. This particularly allows us to include fluxes from the less sensitive $u$ and W4 bands that are below the formal detection limit. Including these sources, the results on the composite SEDs are essentially identical, so for the remainder of the paper we focus on sources with robust detections in all bands of interest.

For
each object, we determine the fluxes at the rest-frame
wavelengths corresponding to the observed photometric bands, and then interpolate between these using piece-wise power laws (linear in log-log space) to
produce a rough SED for each object. We then normalize each object's
SED based on the integrated flux in the rest-frame 8--13 \micron.
This corresponds roughly to the integrated observed flux
between the W3 and W4 {\wise} bands, and so corresponds to the part of the spectrum that has robust flux measurements at the longest wavelengths, where it suffers least from dust attenuation (although, as we note below and discuss in \S~\ref{sec:discussion}, obscuration may still have some effect on the AGN SED at these wavelengths). At each wavelength, the
logarithms of these interpolated and normalized fluxes are then
averaged at each wavelength (eliminating fluxes with a deviation from the mean of $>6\sigma$ to avoid biasing the average due to outliers) and the variance computed.  We thus produce a final composite SED
template for the sources in each subset of QSOs.

\subsection{Comparison of QSO 1 and 2 SEDs}

The average optical--MIR SEDs with the corresponding variance are shown in
Figure~\ref{fig:seds} and tabulated in Table~\ref{tab:seds}. There are clear differences between the SEDs of
the QSO 1s and 2s: The QSO 1 SED is very close to flat in $\nu F_\nu$,
while the QSO 2 SED shows clear features including a break at
$\approx$4000 \AA\ (corresponding to the characteristic break in SEDs
of galaxies) and a redder continuum at $>$4 \micron.

To better understand the differences between these SEDs, we perform
fits to the composite SEDs with a simple multi-component model
consisting of quasar continuum (with dust extinction) and a
host galaxy, with an additional blackbody component at long
wavelengths, as described below. For the quasar SED we use the type 1
quasar template presented in {
six}, and apply reddening using
an extinction curve given by the parameterization of
\citet{fitz99red}. For the host galaxy template we use a linear combination of two empirical galaxy SEDs (E, and Im, corresponding to old and young stellar populations, respectively) derived by \citetalias{asse10agntemp}. We note that the analysis of \citetalias{asse10agntemp} also included an intermediate (Sbc) template, but we find that the E and Im templates alone are sufficient to capture the full variance of the SEDs and so omit the Sbc template for simplicity.
We have also used galaxy SED models computed
from the PEGASE stellar evolution code \citep{fioc97}, with no significant change in the results. We find that the
QSO 1 and 2 SEDs show an excess beyond the {\r06} quasar SED
at $>$8 \micron. We model this excess with a simple blackbody
continuum, with best-fit temperature $\approx$150 K.  {We note that the AGN template from \citetalias{asse10agntemp} does not fit the average SEDs quite as well as the \citetalias{rich06} QSO template, but using that AGN template for the full analysis does not significantly affect our general conclusions (see Section~\ref{sec:a10test} for details).

We perform least-squares fits of this four component model (two galaxy,
absorbed quasar, and blackbody) to the QSO 1 and 2 composite SEDs. To determine the uncertainties in our model fits to these SEDs, we perform bootstrap resampling. Each subset of QSOs is sampled randomly (with replacement) and the composite SEDs are computed and fitted 100 times. The uncertainties in the fit parameters are determine by the variance in the best-fit values among the bootstrap samples.

The fits are shown in Figure~\ref{fig:seds}, and the best-fit parameters
and bootstrap uncertainties are given in Table~\ref{tab:fits}. Note that
in Figure~\ref{fig:seds} the SEDs are normalized to have equal flux in
the {\em unreddened} AGN component, to best illustrate the effects of
reddening. The fits to the two SEDs are remarkably similar, with
comparable relative normalizations of the quasar and galaxy components
(these are within $\approx$20\%) and blackbody temperatures $\approx
150$ K. The main difference is in the reddening of the AGN component.
For the QSO 1s the model fits prefer the lower limit of zero extinction,
while for the QSO 2s the best fit yields $A_V \approx 20$. Another
difference is in the normalizations of the blackbody component, which is
twice as large for the QSO 2s relative to the QSO 1s. This may
indicate that the effects of dust extinction extends well into

\begin{deluxetable*}{lccccccccccc}

\tablewidth{\textwidth}
\tablecaption{Model Fits to Type 1 and 2 QSO Composite SEDs\tnm{a} \label{tab:fits}}
\tabletypesize{\tiny}
\tablehead{
\colhead{Sample}  &
\colhead{$N_{\rm src}$\tnm{b}}  &
\colhead{$\average{z}$} &
\colhead{$\average{L_{\rm [OIII]}}$\tnm{c}} &
\colhead{$F_{\rm E}$} &
\colhead{$F_{\rm Im}$} &
\colhead{$F_{\rm AGN}$\tnm{d}} &
\colhead{$A_V$} &
\colhead{$T_{\rm BB}$ (K)} &
\colhead{Norm$_{\rm BB}$\tnm{e}} &
\colhead{$\log {M_{\rm gal}}$\tnm{f}} &
\colhead{$\log {L_{\rm MIR}^{\rm AGN}}$\tnm{g}}}
\startdata
\\
\cutinhead{Full samples (SDSS+UKIDSS+{\em WISE})}
QSO 1 & 334 & 0.45 & 42.6 & $0.42\pm0.04$ & $0.02\pm0.03$ & $0.56\pm0.01$ & 0 & $153\pm  2$ & $0.038\pm0.004$ & $10.88\pm 0.02$ & $44.85\pm 0.02$ \\
QSO 2 & 406 & 0.44 & 42.6 & $0.43\pm0.03$ & $0.13\pm0.01$ & $0.44\pm0.01$ & $19\pm 1$ & $159\pm  3$ & $0.068\pm0.008$ & $10.96\pm 0.01$ & $44.81\pm 0.02$ \\
\cutinhead{GALEX+SDSS+2MASS+{\em WISE} ($z < 0.2$)}
QSO 1 & 51 & 0.16 & 42.3 & $0.59\pm0.12$ & $0.01\pm0.02$ & $0.39\pm0.03$ & 0 & $139\pm  5$ & $0.050\pm0.012$ & $10.86\pm 0.04$ & $44.45\pm 0.04$ \\
QSO 2 & 76 & 0.15 & 42.3 & $0.54\pm0.06$ & $0.12\pm0.01$ & $0.34\pm0.02$ & $23\pm 1$ & $145\pm  2$ & $0.091\pm0.007$ & $10.78\pm 0.02$ & $44.37\pm 0.03$ \\
\cutinhead{SDSS+UKIDSS+{\em WISE} ($z<0.2$)}
QSO 1 & 28 & 0.17 & 42.3 & $0.51\pm0.18$ & $0.03\pm0.03$ & $0.47\pm0.05$ & 0 & $142\pm  4$ & $0.044\pm0.008$ & $10.72\pm 0.05$ & $44.45\pm 0.06$ \\
QSO 2 & 39 & 0.15 & 42.3 & $0.56\pm0.09$ & $0.11\pm0.02$ & $0.33\pm0.03$ & $22\pm 2$ & $139\pm  3$ & $0.121\pm0.012$ & $10.75\pm 0.03$ & $44.31\pm 0.05$ \\
\cutinhead{[O\textsc{III}] luminosity selection ($\log \loiii/\rm{[erg\;s^{-1}]}$)}
QSO 1 (42.00--42.3) & 67 & 0.31 & 42.1 & $0.53\pm0.08$ & $0.03\pm0.03$ & $0.44\pm0.02$ & 0 & $148\pm  3$ & $0.039\pm0.007$ & $10.78\pm 0.02$ & $44.50\pm 0.03$ \\
QSO 1 (42.3--42.5) & 69 & 0.40 & 42.4 & $0.48\pm0.06$ & $0.00\pm0.01$ & $0.52\pm0.02$ & 0 & $152\pm  2$ & $0.039\pm0.004$ & $10.84\pm 0.03$ & $44.71\pm 0.03$ \\
QSO 1 (42.5--42.7) & 89 & 0.47 & 42.6 & $0.37\pm0.10$ & $0.00\pm0.03$ & $0.63\pm0.02$ & 0 & $150\pm  3$ & $0.056\pm0.008$ & $10.79\pm 0.04$ & $44.88\pm 0.03$ \\
QSO 1 (42.7--43.0) & 74 & 0.52 & 42.8 & $0.38\pm0.09$ & $0.00\pm0.06$ & $0.62\pm0.02$ & 0 & $152\pm  3$ & $0.045\pm0.006$ & $10.95\pm 0.03$ & $45.02\pm 0.03$ \\
QSO 1 (43.0--43.5) & 35 & 0.58 & 43.2 & $0.20\pm0.14$ & $0.18\pm0.13$ & $0.62\pm0.03$ & 0 & $157\pm  4$ & $0.041\pm0.010$ & $11.05\pm 0.07$ & $45.21\pm 0.04$ \\
QSO 2 (42.00--42.3) & 77 & 0.29 & 42.1 & $0.59\pm0.07$ & $0.09\pm0.02$ & $0.32\pm0.02$ & $19\pm 1$ & $151\pm  3$ & $0.072\pm0.010$ & $10.88\pm 0.03$ & $44.45\pm 0.04$ \\
QSO 2 (42.3--42.5) & 70 & 0.40 & 42.4 & $0.46\pm0.08$ & $0.13\pm0.03$ & $0.41\pm0.03$ & $20\pm 2$ & $156\pm  3$ & $0.080\pm0.015$ & $10.89\pm 0.03$ & $44.70\pm 0.04$ \\
QSO 2 (42.5--42.7) & 111 & 0.47 & 42.6 & $0.41\pm0.04$ & $0.14\pm0.02$ & $0.45\pm0.01$ & $18\pm 1$ & $155\pm  3$ & $0.079\pm0.010$ & $10.97\pm 0.02$ & $44.84\pm 0.02$ \\
QSO 2 (42.7--43.0) & 103 & 0.51 & 42.8 & $0.35\pm0.05$ & $0.16\pm0.02$ & $0.49\pm0.02$ & $19\pm 2$ & $162\pm  3$ & $0.069\pm0.010$ & $10.96\pm 0.02$ & $44.92\pm 0.03$ \\
QSO 2 (43.0--43.5) & 45 & 0.58 & 43.2 & $0.26\pm0.06$ & $0.15\pm0.03$ & $0.58\pm0.02$ & $17\pm 2$ & $159\pm  4$ & $0.082\pm0.013$ & $11.11\pm 0.04$ & $45.23\pm 0.04$ \\
\cutinhead{\threefour\ color selection (Vega)}
QSO 1 ($<$0.9) & 31 & 0.39 & 42.4 & $0.59\pm0.07$ & $0.11\pm0.04$ & $0.30\pm0.01$ & 0 & $150\pm  4$ & $0.054\pm0.008$ & $10.86\pm 0.05$ & $44.42\pm 0.05$ \\
QSO 1 (0.9--1.1) & 179 & 0.41 & 42.5 & $0.47\pm0.04$ & $0.03\pm0.03$ & $0.50\pm0.01$ & 0 & $154\pm  2$ & $0.034\pm0.005$ & $10.88\pm 0.02$ & $44.74\pm 0.02$ \\
QSO 1 ($>$1.1) & 124 & 0.50 & 42.7 & $0.22\pm0.09$ & $0.00\pm0.00$ & $0.78\pm0.02$ & 0 & $156\pm  3$ & $0.036\pm0.006$ & $10.70\pm 0.04$ & $45.09\pm 0.02$ \\
QSO 2 ($<$0.5) & 87 & 0.51 & 42.6 & $0.54\pm0.06$ & $0.13\pm0.01$ & $0.33\pm0.02$ & $27\pm 1$ & $153\pm  2$ & $0.097\pm0.018$ & $11.02\pm 0.02$ & $44.72\pm 0.03$ \\
QSO 2 (0.5--0.8) & 79 & 0.44 & 42.6 & $0.49\pm0.05$ & $0.13\pm0.02$ & $0.38\pm0.02$ & $25\pm 1$ & $156\pm  3$ & $0.080\pm0.013$ & $10.97\pm 0.02$ & $44.73\pm 0.04$ \\
QSO 2 (0.8--1.1) & 116 & 0.41 & 42.5 & $0.46\pm0.04$ & $0.12\pm0.01$ & $0.42\pm0.01$ & $18\pm 2$ & $158\pm  4$ & $0.065\pm0.009$ & $10.90\pm 0.02$ & $44.71\pm 0.03$ \\
QSO 2 (1.1--1.4) & 91 & 0.44 & 42.7 & $0.33\pm0.06$ & $0.11\pm0.02$ & $0.56\pm0.01$ & $13\pm 1$ & $155\pm  4$ & $0.070\pm0.013$ & $10.88\pm 0.03$ & $44.92\pm 0.03$ \\
QSO 2 ($>$1.4) & 33 & 0.41 & 42.7 & $0.14\pm0.11$ & $0.08\pm0.04$ & $0.78\pm0.02$ & $17\pm 2$ & $149\pm  7$ & $0.093\pm0.024$ & $10.83\pm 0.05$ & $45.31\pm 0.04$ \\
\cutinhead{Radio selection (RL corresponds to $P_{1.4} > 10^{24}$ W Hz$^{-1}$)}
QSO 1 (RL) & 39 & 0.49 & 42.7 & $0.49\pm0.14$ & $0.00\pm0.08$ & $0.51\pm0.03$ & 0 & $155\pm  3$ & $0.036\pm0.007$ & $11.03\pm 0.06$ & $44.90\pm 0.06$ \\
QSO 1 (RQ) & 66 & 0.50 & 42.7 & $0.42\pm0.08$ & $0.00\pm0.03$ & $0.58\pm0.02$ & 0 & $155\pm  4$ & $0.028\pm0.007$ & $10.99\pm 0.04$ & $44.97\pm 0.04$ \\
QSO 2 (RL) & 107 & 0.49 & 42.7 & $0.36\pm0.07$ & $0.12\pm0.02$ & $0.52\pm0.02$ & $17\pm 3$ & $167\pm  5$ & $0.054\pm0.014$ & $11.06\pm 0.03$ & $45.08\pm 0.03$ \\
QSO 2 (RQ) & 77 & 0.47 & 42.7 & $0.40\pm0.06$ & $0.14\pm0.03$ & $0.45\pm0.02$ & $18\pm 2$ & $162\pm  4$ & $0.063\pm0.012$ & $10.94\pm 0.03$ & $44.84\pm 0.04$ \\
\cutinhead{Full samples (SDSS+UKIDSS+{\em WISE}); \citetalias{asse10agntemp} AGN template}
QSO 1 & 334 & 0.45 & 42.6 & $0.65\pm0.03$ & $0.09\pm0.02$ & $0.26\pm0.01$ & 0 & $143\pm  6$ & $0.017\pm0.006$ & $11.12\pm 0.01$ & $44.84\pm 0.02$ \\
QSO 2 & 406 & 0.44 & 42.6 & $0.58\pm0.03$ & $0.18\pm0.01$ & $0.24\pm0.01$ & $12\pm 1$ & $165\pm  4$ & $0.041\pm0.005$ & $10.93\pm 0.02$ & $44.77\pm 0.02$

\enddata
\tnt{a}{The procedure for creating composite SEDs and details of the individual samples are described in \S~\ref{sec:seds}. Uncertainties in all quantities are statistical only and derived from bootstrap resampling.}
\tnt{b}{Number of unique sources included in each composite SED.}
\tnt{c}{Presented in units of $\log(L_{\rm [OIII]}/{\rm erg\;s^{-1}})$.}
\tnt{d}{The fraction of the total luminosity at 1 \micron\ contributed by the AGN, after correcting the AGN component for dust extinction.}
\tnt{e}{Arbitrary normalization; in each case defined relative to the total SED normalized at 8--13 $\micron$}
\tnt{f}{Presented in units of $\log(M_{\rm gal}/\msun)$.}
\tnt{g}{The average monochromatic $\nu L_\nu$ luminosity at 12 \micron\ (in $\log \left(L_{\rm MIR}^{\rm AGN}/{\rm erg\;s^{-1}}\right)$) of the quasars in this subsample, assuming a contribution from the AGN and corrected for dust extinction using the best template fit to the composite SED. }
\end{deluxetable*}

\noindent the MIR, or an extra contribution due to star formation in QSO 2 hosts; we discuss these possibilities in
\S~\ref{sec:discussion}. The primary conclusion from these model fits is
that the QSO 1 and QSO 2 SEDs can be broadly described by the same
simple model. We use this model in \S~\ref{sec:colors} to make predictions for selection
of these objects based on optical and MIR photometry.

In Figure~\ref{fig:twogalseds}, we also show the composite SEDs including \galex{} and 2MASS data for QSOs with $z<0.2$, as described in \S~\ref{sec:photo}. The SEDs are produced using the method described above (including only sources detected in all photometric bands), and the fits are performed using the same models. For a direct comparison, we also carry out the SDSS+UKIDSS+{\wise} analysis described above, limiting the sources to $z < 0.2$. Fit results for both samples are given in Table~\ref{tab:fits}. The SEDs and model fits including \galex{} and 2MASS are generally consistent with those of the  $z<0.2$ SDSS+UKIDSS+{\wise} comparison SEDs. The most significant difference compared to the full  sample (extending to $z=0.8$) is a smaller $F_{\rm AGN}$ for the quasars at $z<0.2$, as expected for the somewhat lower \loiii\ values of the low-$z$ sources (discussed further in the next section). Finally, we note that the inclusion of the \galex{} data in the QSO 2 SED (Figure~\ref{fig:twogalseds}) reveals an upturn at UV wavelengths. This may be due in part to significant star formation in QSO 2 hosts \citep[e.g.,][]{chen15qsosf}, but may be dominated by scattering from the nuclear source \citep[e.g.,][]{zaka05, zaka06host, obie16qso2}, a component that is not currently included in the SED fits.

\subsection{Dependence of SEDs on \loiii, MIR color, and radio-loudness}

\begin{figure*}
\epsscale{1.15}
\plottwo{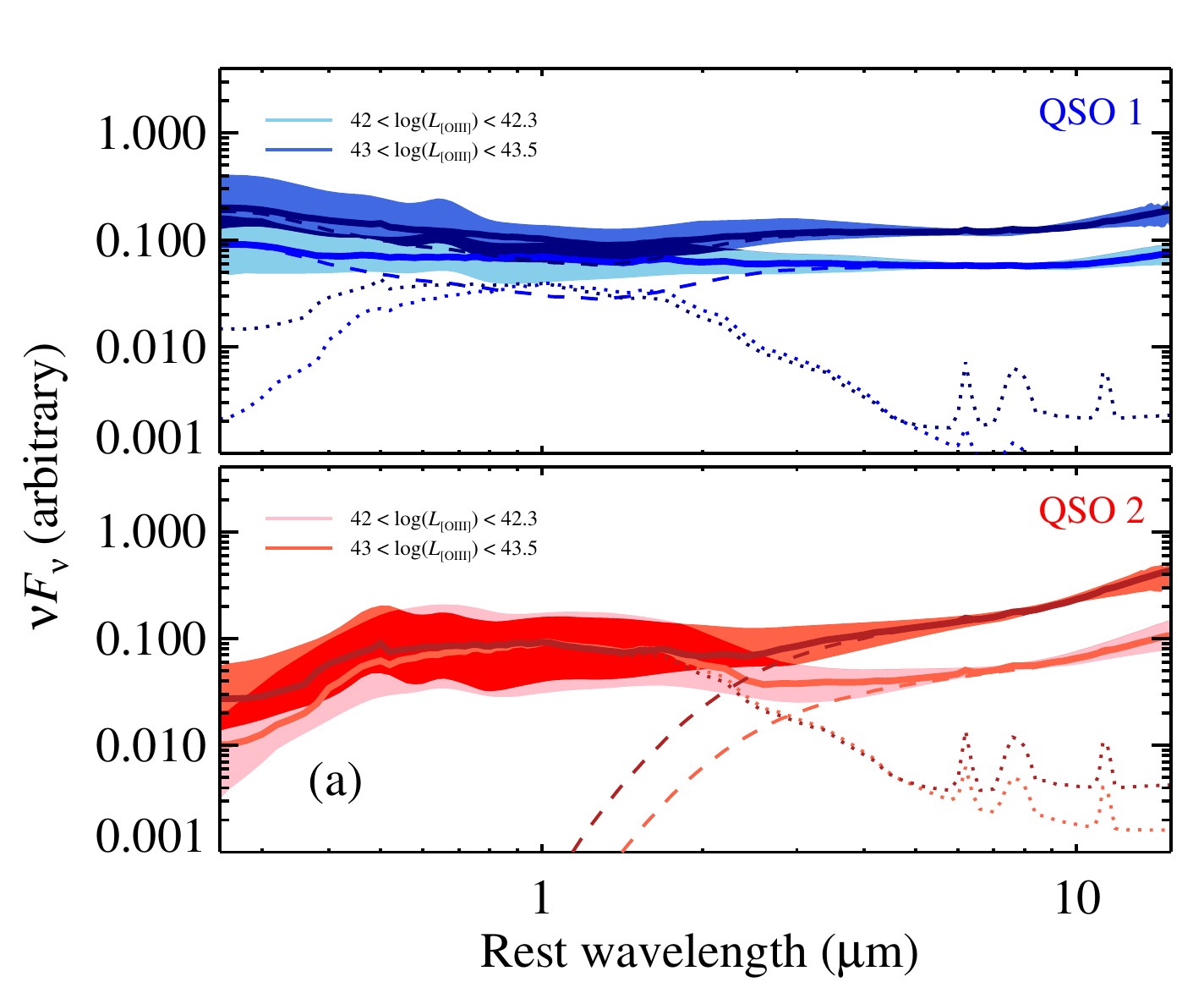}{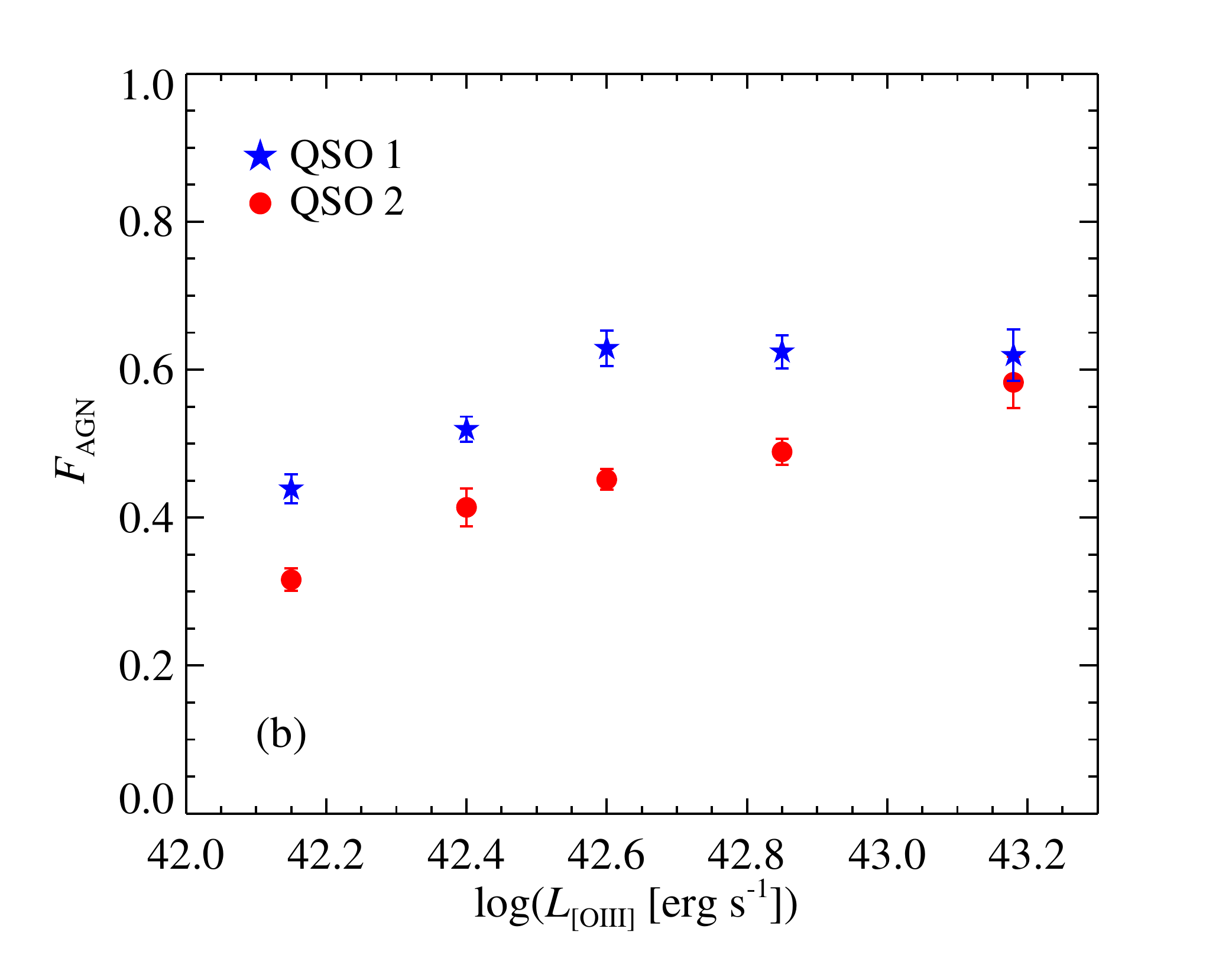}
\caption{Dependence of QSO SEDs on \loiii. (a) Composite SEDs for high- and low-\loiii\ bins for QSO 1s and 2s (produced as described in \S~\ref{sec:loiiirad}) along with model fits. For clarity, the long-wavelength blackbody is included in the AGN component. For each QSO type, the SEDs for the two \loiii\ bins are normalized to have equal flux in the total galaxy component at 1 \micron, to highlight the changing contribution of the AGN relative to the host galaxy. (b) Dependence with \loiii\ of the AGN fraction ($F_{\rm AGN}$), defined as the fraction of the (unabsorbed) AGN component to the total (unabsorbed) luminosity at 1 \micron. For both QSO 1s and 2s, the AGN contributes a relatively larger fraction to the total luminosity as \loiii\ increases. \label{fig:seds_bl} }
\end{figure*} 

 \begin{figure*}
\epsscale{1.15}
\plottwo{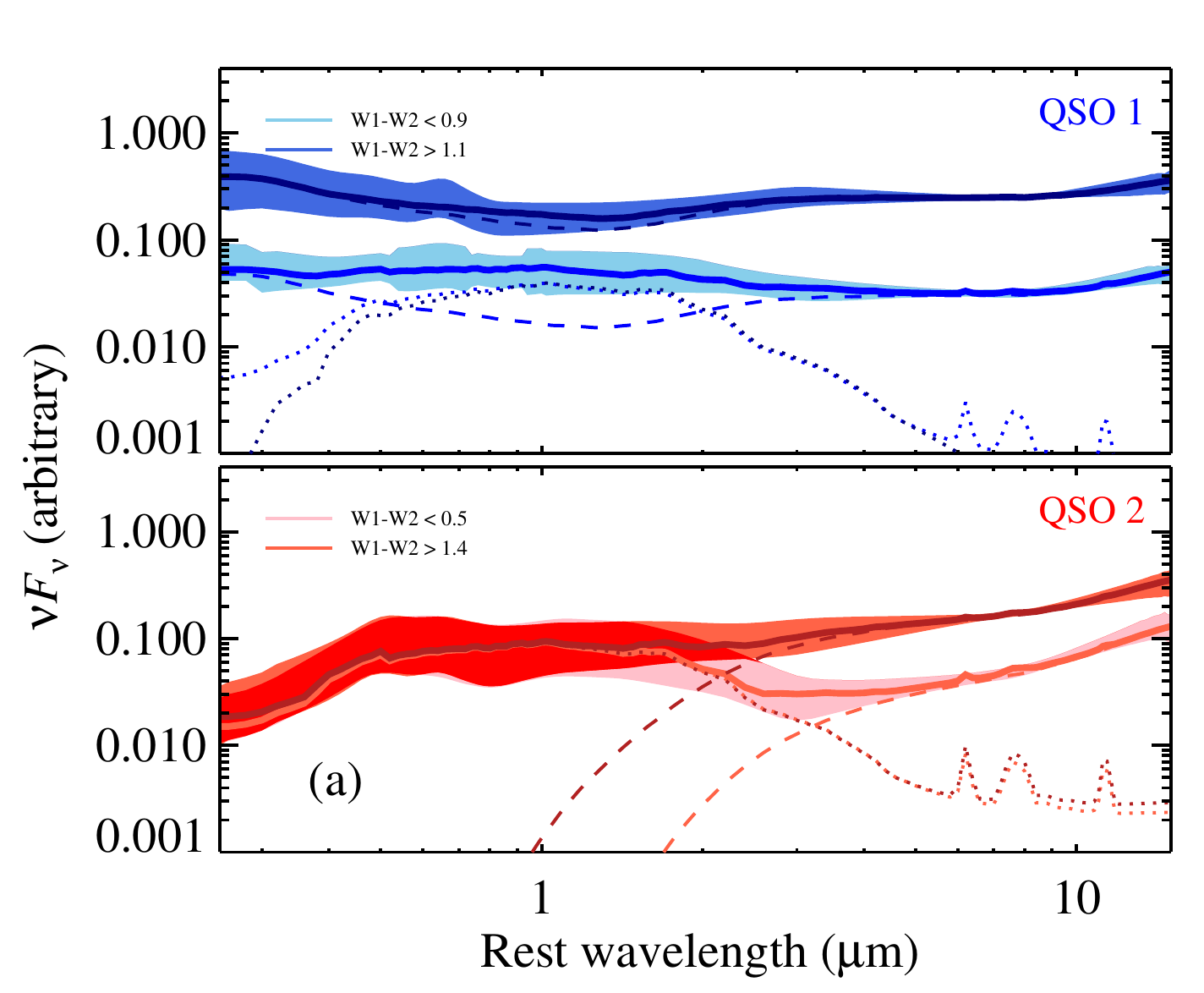}{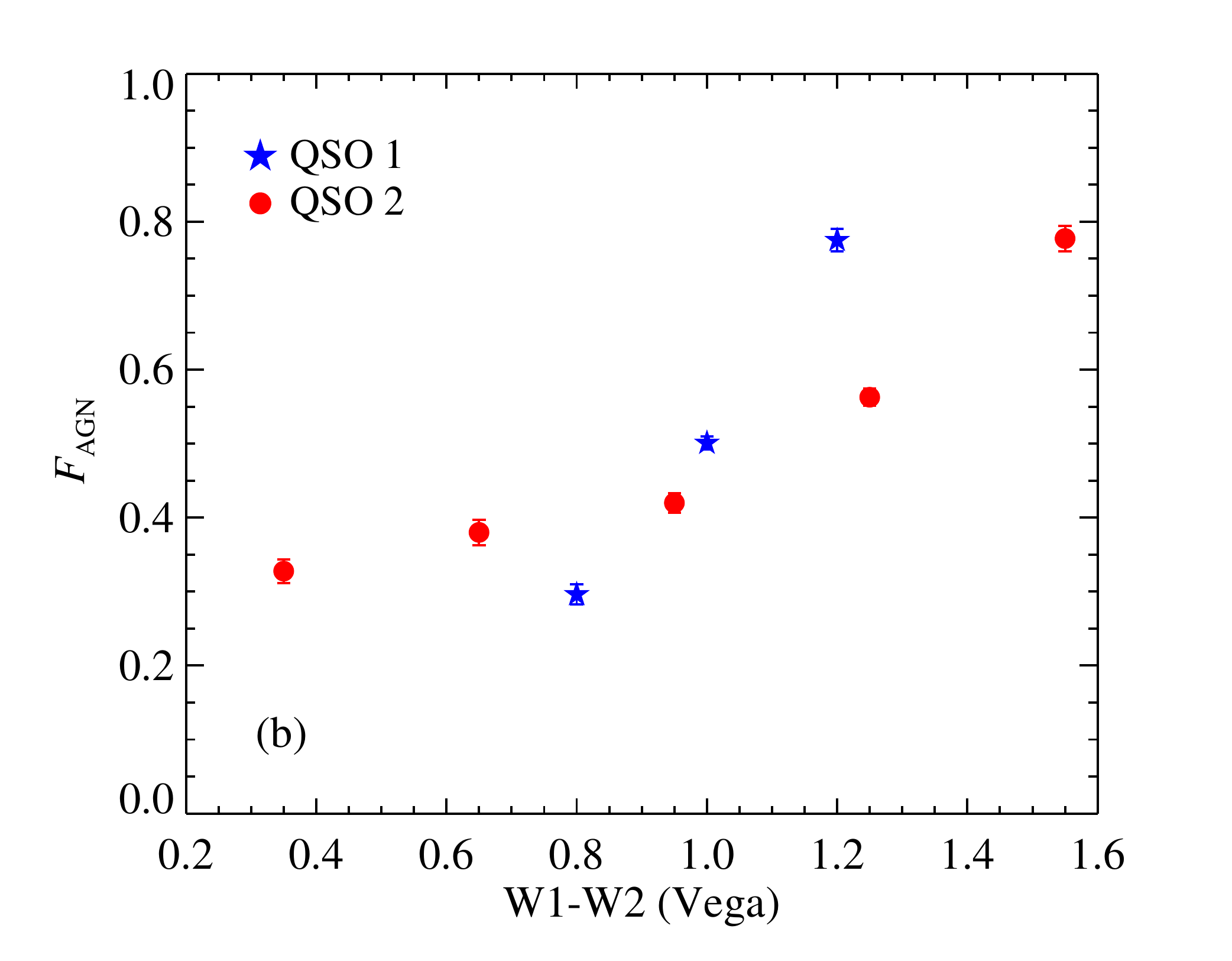}
\caption{Dependence of QSO SEDs on observed \threefour. (a) Composite SEDs for sources with blue and red colors for QSO 1s and 2s (produced as described in \S~\ref{sec:loiiirad}) along with model fits, and normalized as in Fig.~\ref{fig:seds_bl}. (b) Dependence of $F_{\rm AGN}$ with observed \threefour\ color. For both QSO 1s and 2s, the AGN contributes a relatively larger fraction to the total luminosity for redder \threefour, but the dependence is much steeper for QSO 1s, which occupy a smaller overall range in \threefour. \label{fig:seds_w12} }
\end{figure*}

\begin{figure}
\epsscale{1.15}
\plotone{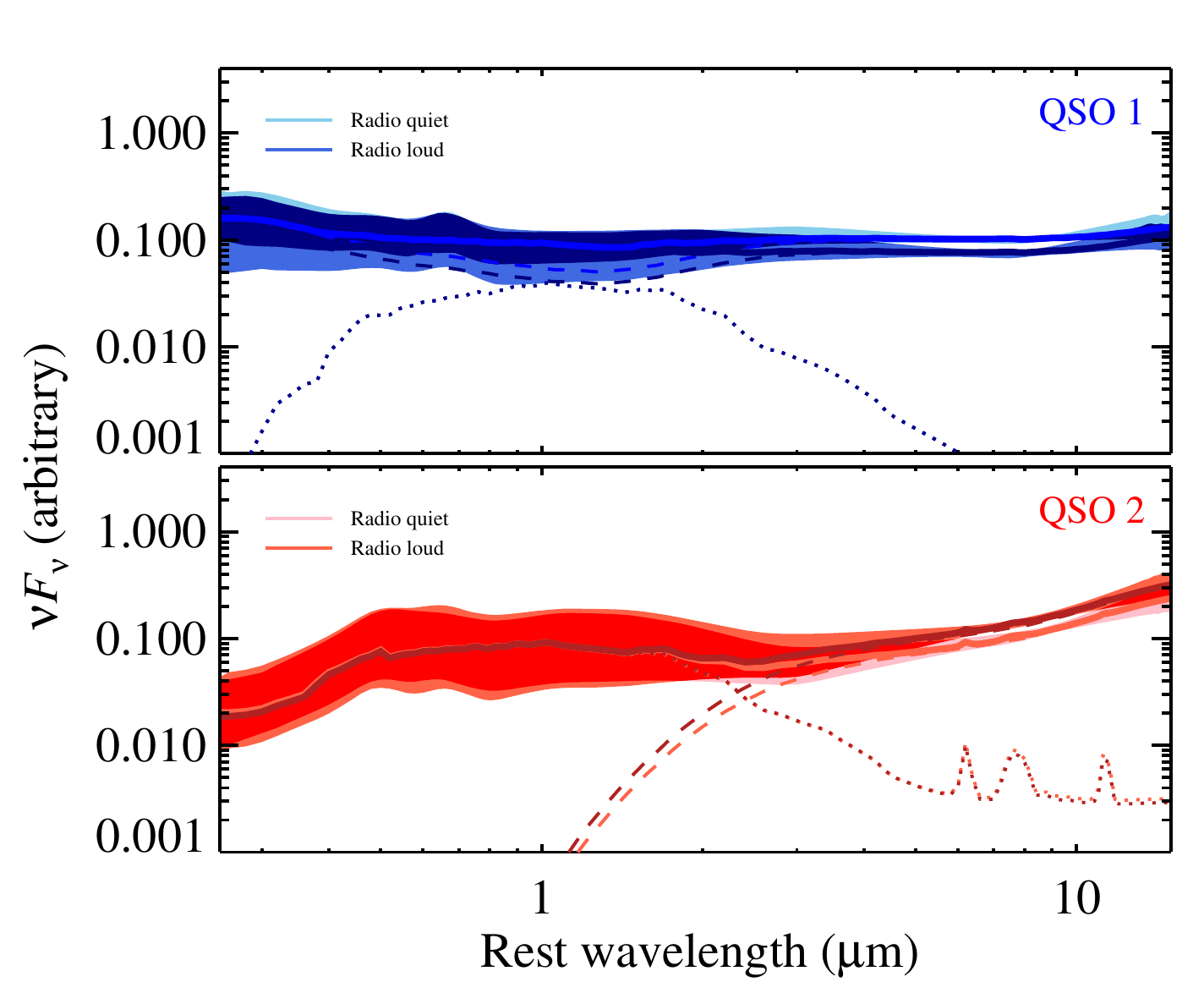}
\caption{Composite SEDs as shown in Figure~\ref{fig:seds_bl}, for the QSOs divided into RL and RQ subsets based on a 1.4 GHz monochromatic luminosity threshold of $P_{\rm 1.4} > 10^{24}$ W Hz$^{-1}$. The composite SEDs of both types of quasar show only a weak dependence of the SEDs on radio-loudness.   \label{fig:seds_rl} }
\end{figure} 

\label{sec:loiiirad}

We next examine the average SEDs as a function of quasar physical and observational
parameters, specifically \loiii, {\wise} color (\threefour), and radio-loudness. We divide the main (SDSS+UKIDSS+{\wise}) samples into five bins of \loiii\ and produce composite SEDs as described in \S~\ref{sec:seds}. The composite SEDs for the highest and lowest \loiii\ bins are shown in Figure~\ref{fig:seds_bl}. In this figure we normalize the SEDs to have equivalent flux in the total {\em galaxy} component at 1 \micron, in order to illustrate the changing contribution of the AGN relative to the galaxy as a function of AGN narrow-line luminosity. In Figure~\ref{fig:seds_bl}(b) we show the ratio of the fluxes of the (unabsorbed) AGN to the total (unabsorbed) SED (AGN and galaxy) at 1 \micron\ ($F_{\rm AGN}$). For every bin of \loiii, $F_{\rm AGN}$ is higher for the QSO 1s than the QSO 2s, perhaps reflecting the photometric selection of the QSO 1 sample, which biases those objects toward sources in which the active nucleus outshines the galaxy in the optical \citep[e.g.,][]{hopk09lowlum}. However, for both QSO 1s and QSO 2s, the contribution of the AGN relative to the galaxy increases modestly with increasing \loiii. 

We also compute composite SEDs as a function of \threefour\ color. Short-wavelength MIR color (as with IRAC [3.6]--[4.5]) can be used as an AGN indicator and thus a proxy for the dominance of the AGN over the host galaxy \citep[e.g.,][]{ster05, hick07abs, ster12wise}.  We again divide the main (SDSS+UKIDSS+{\wise}) samples into bins of \threefour\ and produce composite SEDs. The QSO 1s are divided into fewer bins, due to their smaller range in \threefour\ color. The SEDs for the highest and lowest W1--W2 bins, and the dependence of $F_{\rm AGN}$ on \threefour, are shown in Figure~\ref{fig:seds_w12}. It is immediately clear that \threefour\ color is strongly correlated with $F_{\rm AGN}$, particularly for QSO 1s, and QSO 2s with \threefour\ $>0.8$. For QSO 2s with bluer \threefour, the dependence of $F_{\rm AGN}$ on \threefour\ is weaker, but there is a stronger correlation between the color and $A_V$, as illustrated in Table~\ref{tab:fits}. This suggests that QSO 2s with bluer W1--W2 colors have similar AGN power, but higher obscuration, compared to their redder QSO 2 counterparts, as discussed further in \S~\ref{sec:colors}.

As a final comparison, we create composite SEDs for QSO 1s and 2s divided by radio loudness between RL and RQ. To specifically check the dependence on radio-loudness, the QSO 1 and QSO 2 RQ samples are further selected to match the $z$ and \loiii\ distributions for the corresponding RL subsets.  The composite SEDs are shown in Figure~\ref{fig:seds_rl}, normalized as in Figure~\ref{fig:seds_bl}. It is immediately apparent that the SEDs are comparable, with similar $F_{\rm AGN}$ values; the trends with $F_{\rm AGN}$ and radio loudness are actually opposite for the QSO 1s and QSO 2s. We conclude that there is no strong connection between the optical--MIR SEDs of quasars and the presence or absence of a luminous radio jet.

\begin{figure*}
\epsscale{1.15}
\plotone{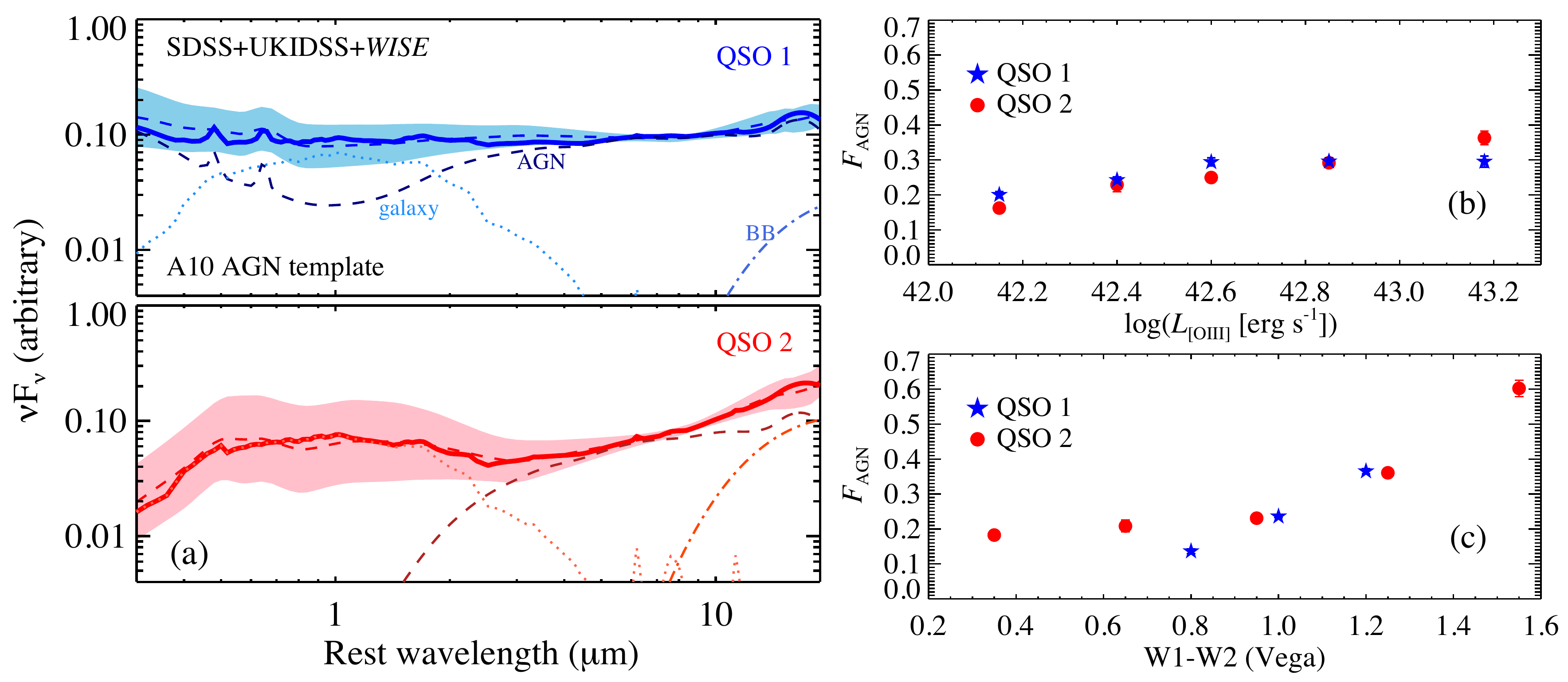}
\caption{Panel (a) shows composite QSO 1 and 2 SEDs shown in Figure~\ref{fig:seds}, with model fits using the \aten\ AGN template, as discussed in Section~\ref{sec:a10test}. The right panels show the variation in $F_{\rm AGN}$ with (b) \loiii\ and (c) \threefour, using the \aten\ AGN template (corresponding to Figures \ref{fig:seds_bl} and \ref{fig:seds_w12}). The model with the \aten\ AGN template broadly fits the composite SEDs, although less well (particularly for QSO 1s) as the \rsix\ template. Using the \aten\ rather than the \rsix\ AGN template returns a lower $F_{\rm AGN}$ with better correspondence between QSO 1s and 2s; the two AGN templates yield similar trends between $F_{\rm AGN}$, \loiii, and \threefour.   \label{fig:a10test} }
\end{figure*} 

\subsection{Fitting results using A10 AGN template}
\label{sec:a10test}

Here we explore the results of our fitting using the AGN template from \citetalias{asse10agntemp}, rather than the template of \citetalias{rich06}. (We use the primary AGN template provided by \citetalias{asse10agntemp}. That paper also presents an AGN template that is derived starting from the \citetalias{rich06} template; the fit results for this template are essentially identical to those described above.) The \citetalias{rich06} template was derived from the SEDs of optically-identified broad-line quasars, while the \citetalias{asse10agntemp} template was computed using a combination of unobscured and obscured AGNs identified using a range of methods (X-ray, MIR, and optical), and over a wider range in luminosity. The results of the fits using the \citetalias{asse10agntemp} are shown in Figure~\ref{fig:a10test} and given in Table~\ref{tab:fits}. The model with the \citetalias{asse10agntemp} AGN template broadly reproduces the shape of the total SEDs, although less well  (particularly for QSO 1s) than the fit with the \citetalias{rich06} AGN template. 

The overall fit results, and the comparison between the fit parameters for QSO 1s and 2s, are similar between the two AGN templates, with the significant exception being the estimate of $F_{\rm AGN}$, which is approximately 25\% when using the \citetalias{asse10agntemp} template, compared to 40--50\% for the \citetalias{rich06} template. This difference is due to the fact that the \citetalias{asse10agntemp} AGN template is weaker in the NIR, so that a stronger galaxy component (corresponding to lower $F_{\rm AGN}$) is needed to fit the observed SEDs (see also \aten{} for a discussion of this effect).  The dependence of $F_{\rm AGN}$ on \loiii\ and W1--W2 is similar between the two templates, with the \citetalias{asse10agntemp} template fitting showing better agreement in $F_{\rm AGN}$ between the two QSO types. The other clear difference between the results is that the {\aten} template is redder in the MIR, requiring a somewhat weaker blackbody component relative to the fits with the {\rsix} AGN template. We note however that for both sets of templates, the QSO 2s require a stronger blackbody component compared to the QSO 1s, reflecting the different shapes of the two SEDs. Using the {\aten} template, the fits are consistent between RL and RQ QSOs, similar to what is found with the {\rsix} template.

We note that the composite SEDs presented here are created using broadly similar techniques to those in \rsix{} and \aten{}, so the reasons for the differences between the templates are not immediately clear. The \aten{} template was constructed from AGN extending to lower luminosities than those for \rsix{}. The fact that our composite SEDs for QSOs are best fitted by the \rsix{} template may imply a dependence of the intrinsic AGN SED with luminosity.  The \aten{} procedure also included a more self-consistent treatment of the host galaxy emission, and \aten{} suggests that the \rsix{} template may be brighter in the NIR due to residual host galaxy contamination. In contrast, it may be possible that the redder colors of the \aten{} template at long wavelengths are due to some cooler dust emission that is either characteristic of lower-luminosity AGN, or originates in star-forming regions in the host galaxy. (The need for a stronger long-wavelength blackbody component in our QSO 2s suggests that this emission may be connected to AGN obscuration.)
Overall, we conclude that the choice of AGN template has some impact on the ultimate fit parameters, so that it is difficult to draw strong conclusions about the absolute strength of the AGN relative to the galaxy luminosity. However, our primary conclusions about the {\em relative} shapes of the QSO 1 and 2 SEDs, and their dependence on luminosity, MIR color, and radio-loudness are independent of the AGN template used.

\section{Photometric selection of luminous obscured quasars}
\label{sec:colors}

A primary challenge for studying the large populations of obscured
quasars detected with {\wise} is selecting these objects based on optical
and MIR photometric data alone, in the absence of spectroscopic or
other multiwavelength indicators. In this section we examine the
observed colors of the SDSS QSO 1 and 2 populations, and explore how
the observed colors of the model SEDs in \S~\ref{sec:seds} vary with
redshift. Here we focus on colors from {\wise} and SDSS, which are among the largest-area MIR and optical surveys with the depth required to detect obscured quasars to $z>1$ \citep[e.g.,][]{asse13wiseagn, dipo14qsoclust, dipo15qsocmb}.

\begin{figure*}
\epsscale{1.15}
\plottwo{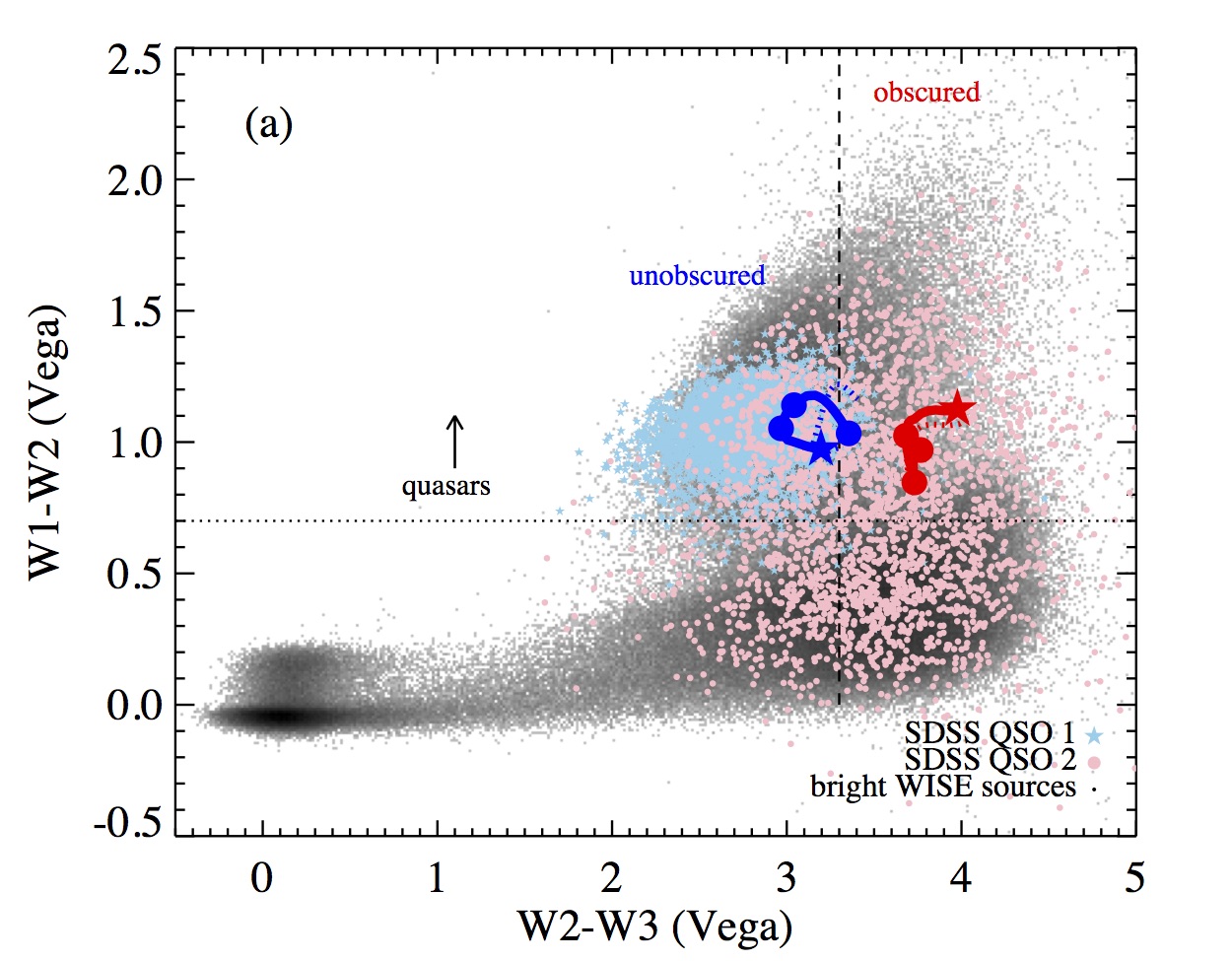}{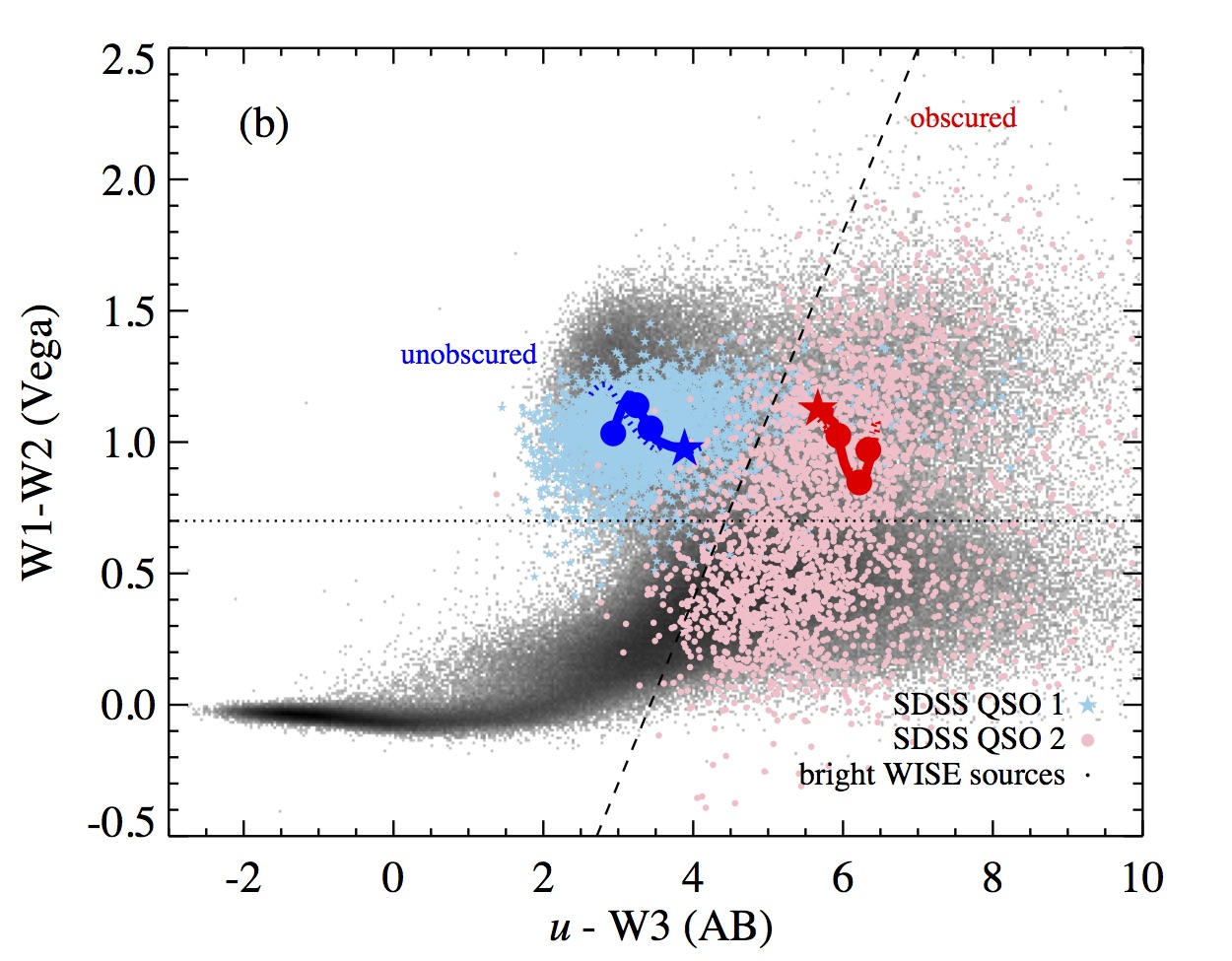}
\caption{{\wise} and SDSS colors of SDSS type 1 and 2 quasars (light blue
  stars and pink circles, respectively), along with redshift tracks
  for composite QSO 1 and 2 SEDs computed as described in
  \S~\ref{sec:seds} and \ref{sec:colors} (red and blue lines,
  respectively). Tracks for the \rsix\ and \aten\ AGN templates are shown as solid and dotted lines, respectively. On the \rsix\ tracks, $z=0$ is indicated by stars and
  $z=0.5$, $z=1$ and $z=2$ by circles. Panel (a) shows {\wise}
  \threefour\ and \fourtwelve\ colors, including a simple \threefour\
  boundary (similar
  to the criteria of \citealt{ster12wise}) to separate quasars from other bright {\wise} sources, and a nominal boundary in
  \fourtwelve\ to separate obscured and unobscured quasars in the sample. The background grayscale shows the distribution of {\wise} sources with W3 detections and W2 $<$ 15.05 (Vega), as applied in recent studies of {\wise} quasar clustering \citep{geac13qsocmb, dipo14qsoclust, dipo15qsocmb, dipo16qsoclust}.
  Panel (b)
  is similar but instead the $x$-axis shows optical to MIR
  ($u$--W3) color, which very effectively separates obscured and
  unobscured quasars that are redder than the \threefour\
  boundary. A putative cut to select obscured quasars, given by $(u-{\rm W3\; [AB]})>1.4({\rm W1-W2\; [Vega]})+3.2$, is shown by the dashed line.  This figure shows that powerful AGNs consistently have red
  MIR colors that can be selected based on a simple cut in
  \threefour\, and that optical-IR colors (and even MIR colors
  themselves at low redshift) can be effective at separating obscured and unobscured
  sources, enabling the selection of large samples of
  obscured quasars from {\wise} and SDSS photometry, although potentially missing many heavily obscured sources with red \threefour\ colors.
 \label{fig:colors}}
\end{figure*}

\begin{figure}
\epsscale{1.15}
\plotone{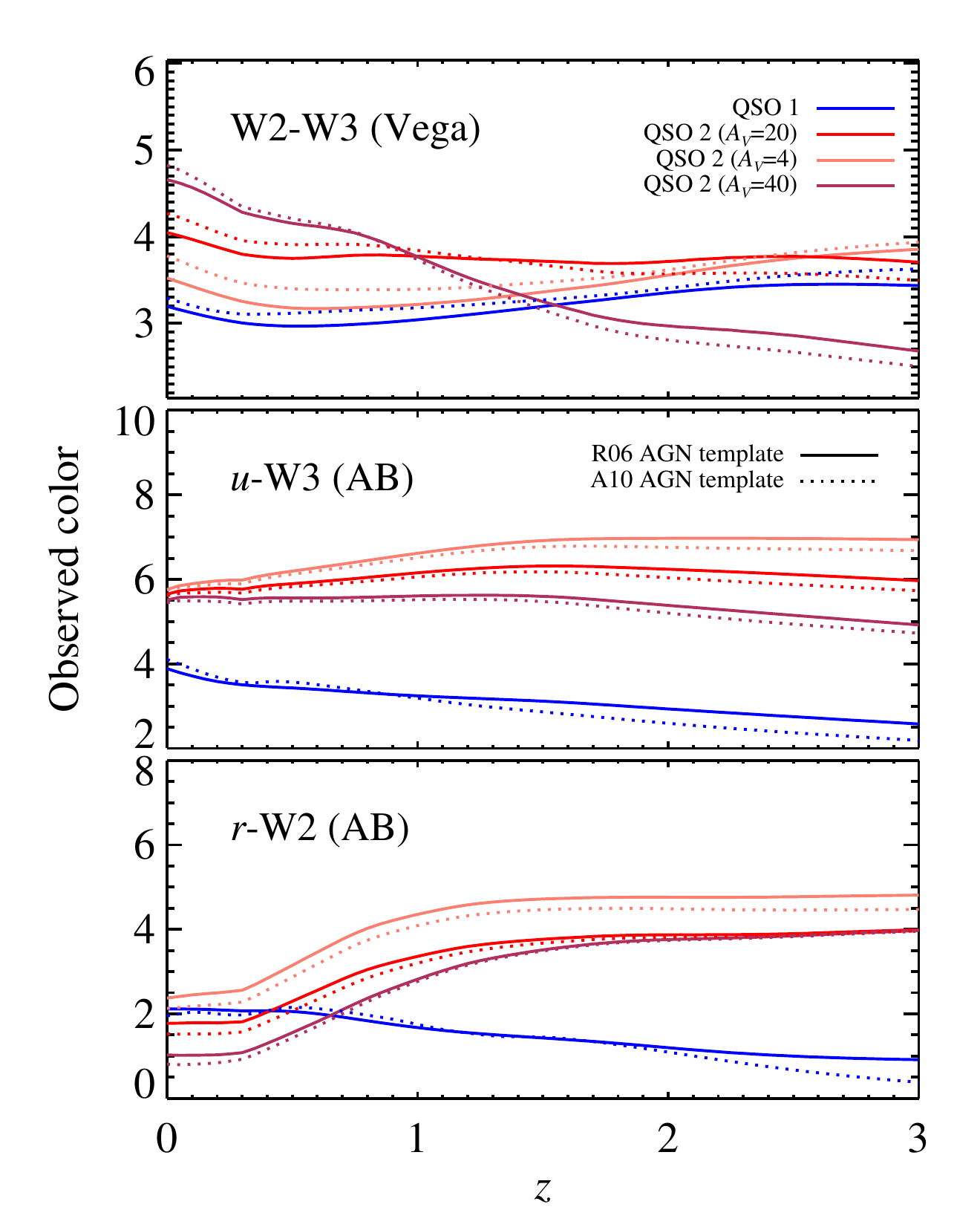}
\caption{Model tracks for MIR and optical-IR colors using the QSO 1
  and 2 composite SED fits, as described in
  \S~\ref{sec:colors}. Tracks are shown for three different levels of
  obscuration in the QSO 2s ($A_V=4$, 20, and 40). The top panel shows
  that \fourtwelve\ color can distinguish between obscured and
  unobscured quasars, but only at low redshift. The middle and bottom
  panels show that optical-IR colors are effective at separating the
  two types of quasars, with clear separation in $u-{\rm W3}$ at all
  redshifts, while $r-{\rm W2}$ distinguishes between them at $z>0.5$.
\label{fig:colz} }
\end{figure}

\subsection{Quasar selection using mid-infrared colors}

A number of simple criteria for selecting AGN using MIR data from {\wise} have
 been proposed \citep[e.g.,][]{jarr11wise, ster12wise,
  mate12xmmwise}, following on similar efforts using \spitzer\ IRAC
and MIPS data \citep{lacy04, ster05, rowa05, donl12irac}. Of the {\wise}
criteria, the simplest is that proposed by \citet{ster12wise},
consisting of a single cut in the \threefour\ color with $\threefour
> 0.8$ (Vega), for sources with a flux threshold of ${\rm W2}<15.05$. This cut picks out objects with red NIR SEDs and
thus separates the (red) hot dust continuum from AGN from the (blue)
approximately Rayleigh-Jeans tail of the stellar continuum, and is
roughly equivalent to the lower bound of the IRAC ``wedge'' selection
presented in \citet{ster05}. To probe to fainter \wise{} fluxes, \citet{asse13wiseagn} proposed a \threefour\ threshold that is dependent on W2 magnitude (with the threshold becoming redder for fainter W2 sources) that achieves $\sim$90\% reliability for ${\rm W2}<17.11$. Other studies have proposed additional
cuts making use of the 12 \micron\ band, specifically the
\fourtwelve\ color \citep{jarr11wise, mate12xmmwise}, to more
efficiently remove contamination from star-forming galaxies. The 22
\micron\ {\wise} band is generally not useful for AGN selection at medium
to high redshift, because of its significantly poorer sensitivity
relative to the three bluer bands \citep[e.g.,][]{mate12xmmwise}.

To compare to these selection criteria, in Figure~\ref{fig:colors}(a) we
show the observed \threefour\ and \fourtwelve\ colors for the SDSS QSO
1s and QSO 2s, along with the distribution for {\wise} sources with W3 detections and W2 $<$ 15.05 (Vega), as selected in recent studies of {\wise} quasars \citet{dipo14qsoclust, dipo15qsocmb}.
Almost all (98\%) of the QSO 1s and roughly half (49\%) of QSO 2s have
red \threefour\ colors that would be selected by a cut of $\threefour > 0.7$ (Vega). In comparison, 18\% of the general population bright {\wise} sources would be selected using this cut. The QSO 2s have a wider range of
      \threefour\ colors and are markedly redder in \fourtwelve,
      reflecting the redder MIR continuum that is apparent in the
      composite SEDs (Figure~\ref{fig:seds}). This suggests the possibility that obscured
      quasars might be selected on the basis of their MIR colors
      alone, with a nominal division around ${\rm W2}-{\rm W3} = 3.3$.
      However, it is immediately clear from Figure~\ref{fig:colors}(a),
      this selection would retain some contamination from unobscured
      quasars. 
      
We further explore how the observed colors of QSO 1s and 2s will
change with redshift, making use of the model fits to the composite
QSO 1 and 2 SEDs presented in \S~\ref{sec:seds}. We calculate the
observed colors of these SEDs as a function of redshift and compare
the colors of the two types of quasars.

In this exercise, we do not simply use the best-fit models for the
composite SEDs at all redshifts. This is because, as a flux-limited
data set, the AllWISE catalog is sensitive to sources
with higher luminosities at high redshift. In particular, we will be
limited by the observed MIR luminosity, which is dominated by the
AGN component, and so more luminous AGN will be selected at higher
redshift.  In contrast, we may expect the overall host galaxy
luminosity to change more slowly with redshift, as quasars are hosted in
systems of similar mass at all redshifts \citep[e.g.,][]{myer07clust1, ross09qsoclust, hick11qsoclust, dipo16qsoclust, dipo17qsoclust}.  Thus at higher
redshift, the {\wise} catalog will preferentially include quasars with a
higher fraction of AGN to host galaxy luminosity; we have confirmed this by fits to composite QSO SEDs in bins of redshift, showing that $F_{\rm AGN}$ rises from $\approx$0.4 at $z=0.3$ to $\approx$0.6 at $z=0.8$. In producing our model curves, we use our best-fit components for the QSO 1s and 2s, but vary the
normalization of the AGN components relative to the host galaxy, so that it follows an extrapolation of the trend observed at lower redshift and reaches $F_{\rm AGN} \approx 0.8$ at $z=2$. For comparison, we also compute the equivalent model curves for the fits using the {\aten} AGN template, with the corresponding variation in $F_{\rm AGN}$ with redshift.

The observed colors are derived by convolving the model SEDs with the
response functions of the SDSS and {\wise} filters. The resulting
\threefour\ and \fourtwelve\ colors for the redshift range $0 < z < 2$
are shown by the tracks in Figure~\ref{fig:colors}(a). (Given the {\wise}
12 \micron\ flux limit, we expect to detect relatively few obscured
quasars at $z>2$.) The tracks using the \rsix\ and \aten\ templates are shown as solid and dotted lines, respectively. Both composite quasar SEDs have red
\threefour\ colors at all redshifts, but we note that many obscured quasars are significantly redder in \fourtwelve, and many are bluer in \threefour, than most of the X-ray AGN used in some previous studies of {\em WISE} color selection  \citep[e.g.,][]{ster12wise, mate12xmmwise}. This may be due to the fact that blue \threefour\ colors are associated with higher levels of obscuration that could absorb X-ray emission strongly enough to drop below survey flux limits \citep[e.g.,][]{ster14nustarwise, lans14nustarqso, lans15nustarqso}. Follow-up spectroscopic and X-ray observations of sources toward the bottom right of {\wise} color-color space indeed confirm their AGN nature \citep{hain14salt, hvid17salt} as well as heavy X-ray obscuration (Yan et al.\ in preparation).

\subsection{Distinguishing between obscured and unobscured quasars}

We have thus established that obscured and unobscured quasars
have characteristic colors in the MIR that can be used to distinguish them from galaxies, and that this selection is able to identify almost all QSO 1s and substantial fraction of QSO 2s.  Here we explore the
possibility of distinguishing between the two types of quasars
using further color criteria.

As clearly shown in Figure~\ref{fig:colors}(a) there is significant
separation between QSO 1s and 2s in the {\wise} colors. At low
redshift, the QSO 2 SED is redder than the QSO 1 SED in \fourtwelve\,
raising the possibility of using {\wise} photometry {\em alone} to both
select quasars and distinguish between obscured and unobscured
subsets.  However, there is overlap in \fourtwelve\ between the
high-$z$ QSO 2s and the QSO 1s, meaning that without further
redshift information, it is difficult to define a single
\fourtwelve\ color cut that efficiently separates QSO 1s and 2s. 

 The top panel of Figure~\ref{fig:colz} shows the redshift tracks in
 \fourtwelve, for the QSO 1 and 2 SEDs. (Model curves for the \aten\ template are plotted as dotted lines, showing similar trends.) For the QSO 2s, we explore the
 effects of obscuration by including models with $A_V = 4$ and $40$ as
 well as $A_V = 20$ which best fits the composite SED.  For
 \fourtwelve, there is clear overlap between high-$z$ QSO 1s and
 low-$z$ QSO 2s, and the color depends significantly on $A_V$. This
 suggests that \fourtwelve\ may have limited utility for separating
 the two quasar types; however, choosing only objects that are very
 blue (${\rm W2}-{\rm W3}<2.5$) or very red (${\rm W2}-{\rm W3}>4$) is likely to yield
 relatively clean, if incomplete, samples of obscured and unobscured
 quasars.

A more effective method to separate obscured and unobscured quasars is
based on their observed optical to MIR color
\citep[e.g.,][]{hick07abs, dono14qsoclust, dipo15qsocmb, dipo16qsoclust, dipo17qsoclust, chen15qsosf}.  By definition, obscured quasars are
extinguished in the optical and UV and thus are dominated by their
host galaxies in those bands, while the MIR is relatively
unaffected by obscuration. For the most effective selection, we first
examine the $u-{\rm W3}$ color, which provides the widest baseline between
SDSS and {\wise} photometric bands (excluding the 22 \micron\ band
due to its poorer sensitivity). The observed $u-{\rm W3}$ color and
the tracks for the composite SEDs are shown in
Figure~\ref{fig:colors}(b) and the middle panel of
Figure~\ref{fig:colz}.  These figures shows a clear separation between
the observed colors of QSO 1s and 2s, and the two model SEDs have
significantly different colors for all redshifts and $A_V$. Thus for
$u$ and 12 \micron\ detections (or $u$ band upper limits), this
provides a very effective method for distinguishing between obscured
and unobscured quasars. We define a selection criterion, shown by the dashed line in \ref{fig:colors}(b), that optimizes the separation between QSO 1s and 2s that can also be selected with W1--W2 $>0.7$. This is given by $(u-{\rm W3\;[AB]})>1.4({\rm W1-W2\;[Vega]})+3.2$. This criterion identifies 93\% of the QSO 1s and 92\% of the QSO 2s in our sample (94\% and 95\%, respectively, of those with W1--W2 $>0.7$).

It is also useful to explore criteria that can select obscured quasars
in bands that are more sensitive than the SDSS $u$ and {\wise} 12
\micron\ bands. One such criterion was proposed by \citet{hick07abs}
using IRAC data, for which a cut of $R-{\rm [4.5]} > 3.1$ (AB) provided
clean separation between IR-selected obscured and unobscured quasars
at $z>0.7$; this criterion has since been used by a number of studies with optical and {\em WISE} data \citep[e.g.,][]{dono14qsoclust, dipo14qsoclust, dipo15qsocmb, dipo16qsoclust, dipo17qsoclust, mend16agnclust}.  Tracks for the corresponding SDSS and {\wise} color
($r-{\rm W2}$) for the model SEDs are shown in the bottom panel of Figure~\ref{fig:colz}.
These observed colors are not able to cleanly separate QSO 1s and 2s at low
redshift (as also observed with \spitzer\ data by \citealt{hick07abs}), but the separation becomes most significant at $z>0.5$ as the
4000 \AA\ break moves into the $r$ band, making this a particularly useful
criterion for selecting obscured quasars if there is some redshift information. (We note that in principle, with estimates of redshift one can identify the level of quasar obscuration based on full model fits to the SEDs, for example as performed in \citealt{hain14salt} and \citealt{hvid17salt}. However, in
some cases only uncertain photometric redshifts are available and full SED fitting can be computationally expensive for large samples, so that there can be significant utility in simple photometric color selection.)

We conclude that optical-IR color selection criteria can be
effective at distinguishing between obscured and unobscured quasars, providing a tool to select large numbers of obscured quasars 
selected based solely on {\wise} and optical photometry.

\section{Discussion}
\label{sec:discussion}

The primary conclusion that emerges from this work is the intrinsic similarity of the optical--MIR SEDs of obscured and unobscured quasars, for which the only marked difference in the SED shape appears to be due to reddening of the AGN component. Further, we find no substantial difference in composite SEDs as a function of radio-loudness, confirming the broad conclusions of previous SED analyses for Type 1 quasars \citep[e.g.,][]{elvi94}. In contrast, there is significant dependence of the observed SED on intrinsic AGN luminosity (parametrized here by \loiii) as the radiative AGN emission becomes more prominent relative to the host galaxy.  This contrast suggests a weak connection between the mechanical power of the quasar (as measured by radio luminosity) and the radiative power, with implications for the launching mechanisms of relativistic jets \citep[e.g.,][]{siko07radio, tadh16radio, pado17agn}.}

The overall uniformity of intrinsic quasar SEDs confirms that the population of luminous AGN may be modeled using relatively simple prescriptions that are independent of obscuration of the central engine or the presence of relativistic jets. We therefore have a robust understanding of photometric selection of obscured quasars based on optical and MIR data, and verify simple color criteria that have been used in previous work \citep[e.g.,][]{hick07abs, hick11qsoclust, dono14qsoclust, dipo14qsoclust, dipo15qsocmb, dipo16qsoclust}. 

We emphasize however that obscured quasar color selection is dependent on redshift, illustrating the need in pure photometric selection for redshift estimates (for example utilizing galactic features in the rest-frame optical spectrum) which can be obtained using template fitting \citep[e.g.,][]{hick07abs,chun14sed, hain14salt} or machine learning techniques \citep[e.g.,][]{brod06, geac12map}. We also note that simple MIR color techniques are likely to miss the most heavily obscured AGN, which have MIR colors consistent with star-forming galaxies at similar redshift \citep[e.g.,][]{yuan16qso2, hain16wise}. 

A related point is that the observed SEDs of type 1 and 2 quasars differ even well into the rest-frame MIR, indicating that dust affects even the longest wavelengths probed by {\wise}. This highlights the challenge of fully correcting for these effects, and selecting samples of obscured and unobscured quasars matched in intrinsic luminosity based on photometry alone \citep[e.g.,][]{hick07abs}. The extinction of the AGN component in the observed {\wise} bands may also contribute to the larger flux of the long-wavelength blackbody component that we obtain for the QSO 2s compared to the QSO 1s; a more complete correction for this extinction could in principle yield more consistent contributions of the AGN relative to the galaxy in the two types of QSOs. Alternatively, the stronger long-wavelength emission in QSO 2s could be due to cooler dust heated by star formation, which may be connected to obscuration in quasars \citep[e.g.,][]{page04submm, chen15qsosf}.  Finally, we note that the observed SED differences raise the possibility that quasar MIR SEDs, and inferred intrinsic luminosities, are influenced by orientation effects due to anisotropic absorption and emission \citep[e.g.,][]{hoen11agnsed, podi15qso}. This anisotropy is a natural product of some models of a circumnuclear torus \citep[e.g.,][]{nenk08torus, netz15agn, stal16torus}, while the simple extinction prescription adopted here is more appropriate for a foreground ``screen'' of dust on larger scales in the host galaxy \citep[e.g.,][]{goul12comp, chen15qsosf, glik15qsomerge}.

Based on the SED fits, we can estimate the stellar masses corresponding to the ``average'' host galaxy components. We use the average total luminosities of the quasar in each sample at rest-frame $K$ band (2.1 \micron) and compute the average galaxy luminosity from the components of the best-fit SED model. We then convert to an approximate galaxy stellar mass assuming a $K$-band mass-to-light ratio of 3.1 \citep{cour14galmass}. The resulting galaxy masses are listed in Table~\ref{tab:fits}. For reference, we similarly compute the unabsorbed rest-frame 12 \micron\ luminosity $L_{\rm MIR}^{\rm AGN}$ for each quasar subset, also listed in Table~\ref{tab:fits}. For each quantity, the statistical uncertainties are derived from bootstrap resampling.

The average QSO 1 and 2 SEDs correspond to  similar galaxy stellar masses of $\log{M_{\rm gal}/\msun} \approx 11$. The precise masses depend on the choice of AGN template and so are difficult to estimate precisely; for the {\rsix} AGN template, the resulting QSO 2s galaxy masses are modestly larger than those for the QSO 1s, consistent with the latest measurements of {\wise}-selected quasar clustering \citep[e.g.,][]{dipo16qsoclust, dipo17qsoclust} that indicate that QSO 2s lie in more massive systems. However, for the {\aten} template, the QSO 1s are correspondingly more massive, in conflict with the simplest interpretation of the clustering results. These results highlight the challenges in characterizing absolute galaxy masses for AGN hosts from analyses of broad-band SEDs, and motivate continued study of the shape of the intrinsic AGN SED and its variation with AGN properties.

Independent of the uncertainty on the stellar masses, the dependence of the SED shape for each type of QSO on luminosity (Figure~\ref{fig:seds_bl}) tells us something about the Eddington ratio distribution for luminous quasars. If the luminosity, and thus mass of a galaxy correlates broadly with BH mass (due to BH-galaxy correlations; for a review see \citealt{korm13bh}), then we can interpret $F_{\rm AGN}$ as a rough proxy for $L_{\rm AGN}/M_{\rm BH}$, and thus the Eddington ratio. We find that $F_{\rm AGN}$ increases with \loiii, but only modestly (a factor of 10 in \loiii\ corresponds to a factor of $\sim$2 in $F_{\rm AGN}$. This therefore indicate that quasars of higher luminosity have higher BH masses {\em and} higher Eddington ratios, consistent with a scenario in which AGN with a given BH mass occupy a wide range in accretion rates \citep[e.g.,][]{hopk09bulb, aird12agn, hick14agnsf, jone16agn, jone17agn}.

Finally, we use our proposed selection cuts to estimate the size of the population of  quasars that can be identified using {\em WISE} photometry, based on our previous studies of {\em WISE} quasars
\citep[e.g.,][]{geac13qsocmb,dipo14qsoclust, dipo15qsocmb} and the well-studied IRAC-selected quasar sample of
\citet{hick07abs} in the \bootes\ field. Adopting the color criteria of ${\rm W1}-{\rm W2} > 0.8$ (Vega) and W2 $< 15.05$ (as per \citealt{ster12wise}) and $L_{\rm bol}
\gtrsim\,10^{45}$ \ergs\ (``quasar'' luminosities), we obtain a
population of $\sim40$ unobscured and $\sim$20 obscured quasars per
deg$^2$, with a significant tail out to $z\gtrsim 2$. (While the observed obscured population is smaller, previous studies suggest that after accounting for obscuration and flux limits, the intrinsic populations of obscured and unobscured IR-selected quasars are close to equal in number; e.g., \citealt{hick07abs, asse15wiseqso}.) Over the whole
sky, this corresponds to $\sim$2.5 million {\em WISE}-selected
quasars, with $\sim$900,000 sources that would be selected as obscured based on a color criterion of $r - {\rm W2} > 3.1$ (AB), as employed in some previous work \citep{dipo14qsoclust, dipo15qsocmb, dipo16qsoclust, dipo17qsoclust}.  These estimates highlight the power of {\em WISE} and SDSS photometric selection in identifying large samples of obscured quasars that are dramatically expanding our understanding of the population of rapidly growing black holes.  

\nocite{tayl05topcat}

\begin{acknowledgements}

R.C.H.\ acknowledges support from the National Science Foundation through AST grant numbers 1211112 and 1515364 and CAREER grant 1554584, from NASA through grants NNX16AN48G and NNX15AU32H, and from an Alfred P.\ Sloan Research Fellowship. A.D.M.\ acknowledges support from the National Science Foundation through AST grant number 1515404. R.C.H., A.D.M, K.N.H., and M.A.D.\ acknowledge support from NASA for ADAP grant numbers NNX12AE38G and NNX15AP24G.

This publication makes use of data products from the Wide-field Infrared Survey Explorer, which is a joint project of the University of California, Los Angeles, and the Jet Propulsion Laboratory/California Institute of Technology, funded by the National Aeronautics and Space Administration. This work is based in part on data obtained as part of the UKIRT Infrared Deep Sky Survey.

This publication makes use of data products from the Two Micron All Sky Survey, which is a joint project of the University of Massachusetts and the Infrared Processing and Analysis Center/California Institute of Technology, funded by the National Aeronautics and Space Administration and the National Science Foundation.

Funding for SDSS-III has been provided by the Alfred P. Sloan Foundation, the Participating Institutions, the National Science Foundation, and the U.S. Department of Energy Office of Science. The SDSS-III web site is http://www.sdss3.org/.

SDSS-III is managed by the Astrophysical Research Consortium for the Participating Institutions of the SDSS-III Collaboration including the University of Arizona, the Brazilian Participation Group, Brookhaven National Laboratory, Carnegie Mellon University, University of Florida, the French Participation Group, the German Participation Group, Harvard University, the Instituto de Astrofisica de Canarias, the Michigan State/Notre Dame/JINA Participation Group, Johns Hopkins University, Lawrence Berkeley National Laboratory, Max Planck Institute for Astrophysics, Max Planck Institute for Extraterrestrial Physics, New Mexico State University, New York University, Ohio State University, Pennsylvania State University, University of Portsmouth, Princeton University, the Spanish Participation Group, University of Tokyo, University of Utah, Vanderbilt University, University of Virginia, University of Washington, and Yale University.

\end{acknowledgements}

\end{document}